\documentclass[final,5p,times,twocolumn]{elsarticle}

\usepackage{manuscript/src/mypreamble}

\newacronym{TES}{TES}{Thermal Energy Storage}
\newacronym{BAT}{BAT}{Battery}
\newacronym{BLG}{BLG}{Building block}

\begin{document}

    \begin{frontmatter}

\title{Can occupant behaviors affect urban energy planning? \\Distributed stochastic optimization for energy communities}

\author[TUeaddress,DTUaddress]{Julien Leprince \corref{mycorrespondingauthor}}
\address[TUeaddress]{Technical University of Eindhoven, 5 Groene Loper, Eindhoven 5600 MB, The Netherlands}
\cortext[mycorrespondingauthor]{Corresponding author}
\ead{j.j.leprince@tue.nl}

\author[DTUaddress]{Amos Schledorn}
\address[DTUaddress]{Technical University of Denmark, Building 303B Matematiktorvet, Lyngby 2800, Denmark}
\ead{amosc@dtu.dk}

\author[Twente]{Daniela Guericke}
\address[Twente]{University of Twente, Drienerlolaan 5, 7522 NB Enschede, The Netherlands}
\ead{d.guericke@utwente.nl}

\author[DTUaddress]{Dominik Franjo Dominkovic}
\ead{dodo@dtu.dk}

\author[DTUaddress]{Henrik Madsen}
\ead{hmad@dtu.dk}

\author[TUeaddress]{Wim Zeiler}
\ead{w.zeiler@tue.nl}

\begin{abstract}
To meet carbon emission reduction goals in line with the Paris agreement, planning resilient and sustainable energy systems has never been more important.
In the building sector, particularly, strategic urban energy planning engenders large optimization problems across multiple spatiotemporal scales leading to necessary system scope simplifications.
This has resulted in disconnected system scales, namely, building occupants (bottom layer) and smart-city energy networks (top layer).
This paper intends on bridging these disjointed scales to secure both resilient and more energy-efficient urban planning thanks to a holistic approach.
The intent is to assess the aggregated impact of user behavior stochasticities on optimal urban energy planning.
To this end, a stochastic energy community sizing and operation problem is designed, encompassing multi-level utilities founded on energy hub concepts for improved energy and carbon emission efficiencies.
To secure the scalability of our approach, an organic spatial problem distribution suitable for field deployment is put forth, validated by a proof of concept.
Uncertainty factors affecting urban energy planning are particularly examined through a local sensitivity analysis, namely, economic, climate, and occupant-behavior uncertainties.
Founded on historical measurements a typical Dutch energy community composed of 41 residential buildings is designed.
Results disclose a fast-converging distributed stochastic problem, where boilers are showcased as the preferred heating utility, and distributed renewable energy and storage systems were identified as unprofitable for the community.
Occupant behavior was particularly exposed as the leading uncertainty factor impacting energy community planning. 
This demonstrates the relevance and value of our approach in connecting occupants to cities for improved, and more resilient, urban energy planning strategies.
\end{abstract}


\end{frontmatter}

    

    

\renewcommand\nomgroup[1]{%
  \item[\bfseries
  \ifstrequal{#1}{P}{Superscripts}{%
  \ifstrequal{#1}{B}{Subscripts}{%
  \ifstrequal{#1}{S}{Symbols}{}}}%
]}

\nomenclature[S]{$E$}{electric energy [kWh]}
\nomenclature[S]{$\boldsymbol{E}$}{electric energy vector [kWh]}
\nomenclature[S]{$\boldsymbol{\dot{E}}$}{electric power [W]}

\nomenclature[S]{$Q$}{thermal heat energy [kWh]}
\nomenclature[S]{$\boldsymbol{Q}$}{thermal heat energy vector [kWh]}
\nomenclature[S]{$\boldsymbol{\dot{Q}}$}{thermal heat power [W]}

\nomenclature[S]{$T$}{temperature [K]}
\nomenclature[S]{$\boldsymbol{T}$}{temperature vector [K]}
\nomenclature[S]{$V$}{volume [L]}
\nomenclature[S]{$C$}{capacity [J] or thermal capacity [W/K]}
\nomenclature[S]{$R$}{thermal resistance [K/W]}
\nomenclature[S]{$r$}{interest rate [-]}
\nomenclature[S]{$s$}{slack variable}
\nomenclature[S]{$\boldsymbol{s}$}{slack variable vector}
\nomenclature[S]{$U$}{thermal  [W/K] or [W/m$^2$K]}

\nomenclature[S]{$\alpha$}{constant parameter [-]}
\nomenclature[S]{$\eta$}{efficiency [-]}
\nomenclature[S]{$\sigma$}{self discharge rate [-]}
\nomenclature[S]{$\chi$}{existence (boolean variable) [-]}
\nomenclature[S]{$\tau$}{lifetime [years]}
\nomenclature[S]{$\gamma$}{power coefficient [1/h]}
\nomenclature[S]{$\Omega$}{set of scenarios}
\nomenclature[S]{$\omega$}{scenario index}
\nomenclature[S]{$\pi$}{scenario realization probability}
\nomenclature[S]{$\lambda$}{random variable}

\nomenclature[S]{$t_s$}{time step [hour]}
\nomenclature[S]{H}{time horizon [hour]}

\nomenclature[S]{$O$}{objective term [\euro]}
\nomenclature[S]{$p$}{price or cost [\euro/kWh]}
\nomenclature[S]{$\boldsymbol{p}$}{price or cost vector [\euro/kWh]}

\nomenclature[S]{$\mathbb{B}$}{building set [-]}
\nomenclature[S]{$\mathbb{C}$}{community system set [-]}

\nomenclature[B]{b}{building}
\nomenclature[B]{BOL}{boiler}
\nomenclature[B]{BAT}{battery}
\nomenclature[B]{HP}{heat pump}
\nomenclature[B]{HYD}{hydrogen storage}
\nomenclature[B]{EL}{electrolyzer}
\nomenclature[B]{FC}{fuel cell}
\nomenclature[B]{TES}{thermal energy storage}
\nomenclature[B]{SP}{space heating}
\nomenclature[B]{PV}{photovoltaic panel}
\nomenclature[B]{STC}{solar thermal collector}
\nomenclature[B]{COM}{community}
\nomenclature[B]{HV}{high-voltage}
\nomenclature[B]{MV}{medium-voltage}
\nomenclature[B]{LV}{low-voltage}
\nomenclature[B]{w}{window}
\nomenclature[B]{ch}{charge}
\nomenclature[B]{dch}{discharge}
\nomenclature[B]{gas}{gas}
\nomenclature[B]{U}{unit}
\nomenclature[B]{lvl}{levelized costs}
\nomenclature[B]{nom}{nominal}
\nomenclature[B]{T}{temperature}

\nomenclature[P]{$d$}{decision variable}
\nomenclature[P]{$T$}{vector transpose}
\nomenclature[P]{$inv$}{investment}
\nomenclature[P]{$opr$}{operation}
\nomenclature[P]{$eco$}{economic circumstances}
\nomenclature[P]{$occ$}{occupant behavior}
\nomenclature[P]{$clim$}{climate conditions}
\nomenclature[P]{$slk$}{slack}
\nomenclature[P]{$tot$}{total}
\nomenclature[P]{base}{baseline}
\nomenclature[P]{amb}{ambient}
\nomenclature[P]{dist}{distribution}
\nomenclature[P]{sol}{solar}
\nomenclature[P]{i}{inside}
\nomenclature[P]{e}{envelope}
\nomenclature[P]{m}{medium}
\nomenclature[P]{s}{sensor}
\nomenclature[P]{h}{heater}

\printnomenclature

    \section{Introduction}
Shifting our energy systems to resilient, and sustainable processes has never been more important than today. To tackle the global climate crisis and meet net-zero targets set by the European Green Deal \cite{doi/10.2775/19001}, in line with the Paris agreement \cite{agreement2015paris}, countries around the world urgently need to decarbonize their economies by 2050. This requires them to simultaneously reduce their current energy demand while significantly increasing the penetration of renewable energy sources in decentralized energy systems \cite{perry2008integrating}. 
Recent statistics reveal the building sector as the largest global energy-related $\text{CO}_2$ emission contributor \cite{iea_2020}, consequently placing it as the primary policy target of multiple regions of the globe \cite{zhang2021policy,lu2012effectiveness,economidou2020review,d2017towards}.
A reliable integration of decentralized energy generation systems into the grid, such as photovoltaics, wind energy converters, geothermal heat pumps, or biomass-driven combined heat and power \cite{perry2008integrating}, is, however, challenging due to the variability of weather-dependent sources \cite{eltigani2015challenges}. Couplings to energy storage utilities with robust and flexible control strategies are subsequently required to ensure energy demand and supply meet.
To increase the reliability of renewable and sustainable energy systems, smart grid technologies and demand-side management approaches have been exploited over the last decades to profit from available energies more efficiently. Thereby, peaks in electricity demand can be shifted to periods where energy from intermittent renewable sources is available \cite{palensky2011demand}.

The concept of energy hubs and communities emerged from these ideas, to create autonomous areas optimally supplied with multiple energy sources. 
Energy communities are defined by the European Commission as a “legal entity which is effectively controlled by local shareholders or members, generally value rather than profit-driven, involved in distributed generation and in performing activities of a distribution system operator, supplier or aggregator at a local level” \cite{european2017proposal}.
They form a combination of distribution, conversion, and storage technologies controlled to supply communal consumers of energy. Such consumers represent individual households or apartments but also large building complexes or district facilities.
Typical energy communities extend over the urban energy system as districts. They integrate renewables such as photovoltaics, wind turbines, solar thermal collectors, or hybrid collectors with buildings and are connected to local and regional scale distribution technologies such as smart (micro-)grids and district heating \& cooling networks \cite{kim2019techno,ur2019towards,gjorgievski2021social}.
The design of such communities is not a trivial task and necessitates computational methods gathering multiple energy sources and technologies while optimizing urban to user-level energy flows \cite{orehounig2015integration}.
If done correctly, however, the pooling of communal resources into energy planning has demonstrated significant energetic and economical gains \cite{mohammadi2018optimal}.
For instance, Orehouning et al. \cite{orehounig2015integration} showed that combining energy supply and local energy storage systems together lowered energy demand peaks on the electrical grid and reduced the overall consumption of the neighborhood.
Maroufmashat et al. \cite{maroufmashat2015modeling} demonstrated that developing synergies between up to three energy hubs resulted in significant economic and carbon emission reduction gains, i.e. 11\% to 29\%, as well as a 13\% reduction in natural gas consumption.

The inherent challenge in planning and controlling such systems stems from the stochastic processes driving its three founding pillars, i.e., (\textit{i}) investment strategies, (\textit{ii}) renewable productions, and (\textit{iii}) energy demands. The sources of these uncertainties can be attributed to either of these distinct phenomenons, i.e., (\textit{i}) economic circumstances, (\textit{ii}) climate conditions, and (\textit{iii}) building occupant behaviors.
Economic and climate-related uncertainties are important factors commonly considered in the design of urban energy systems \cite{moret2017strategic}. These provide a uniform setting for the planning of energy districts and have been amply investigated in recent years \cite{haurie2012modeling, santos2016methodology,usher2013expert}.
Occupant behavior, on the other hand, is a notoriously heterogeneous constituent of building energy systems. Driven by multiple contextual, sociological, or psychological factors, they are exceedingly tedious to characterize \cite{hu2020systematic} and have consequently become the leading source of uncertainty in predicting building energy use \cite{hoes2009user,yan2015occupant} contributing to the so-called building performance gap \cite{paone2018impact}.
These behaviors commonly include interactions with thermostats, plug-in appliances, operable lights, windows, or blinds. The control of window blinds by occupants may be motivated by factors such as the desire to either secure privacy or maintain view or a sense of connection to the outdoors for instance \cite{yan2015occupant}.

Under these circumstances, urban energy planners typically leverage energy demand measurements induced from occupant behavior to exploit samples of identified behaviors in the design phase \cite{virote2012stochastic}. It becomes, however, increasingly precarious to develop systems resilient to behavioral variations that are likely to come from either demographic or behavioral transformations \cite{happle2018review}. Subsequently, there exists, to this date, no study examining the impact of varying behavioral groups on strategic urban energy planning.

This shortcoming is typically due to the scale and difference in modeled details between building to room-level energy management problems and urban energy planning ones \cite{aliabadi2021coordination}.
Energy planning problems at the neighborhood, city, or country scale typically need to reduce the encompassed dimensionality through spatial and temporal aggregations to render resulting optimization problems computationally tractable. For example, the planning of a residential neighborhood would consider both a typical, representative, year of operation, to reduce the temporal dimension (horizon) of the problem, as well as aggregated energy demands from clusters of buildings or apartments to simultaneously downscale its spatial granularity \cite{stadler2016model,stadler2018contribution}.
Yet, these necessary simplifications deprive planners from exploiting the full extent of available synergies between prosumers of energy communities. Activating untapped energy flexibility potentials such as demand-side management in the planning phase could significantly improve system efficiency and reduce planning costs.
The question of relevant scale identification in urban energy planning is in fact, not a new one. Cajot et al. \cite{cajot2017obstacles} stated that it should be regarded rather as an open question, for future research to provide planners and decision-makers with rigorous and systematic tools necessary to quantify the gains and losses of different boundaries.

\subsection{Motivation}

In this context, it becomes clear that unifying occupant and building-level information to the urban energy infrastructure can uncover significant reductions in carbon emission and energy demand while allowing the design of systems resilient to the intrinsic uncertainty induced by occupants.
To bring these typically disconnected scales together, we consider a neighborhood system taking the role of an energy community. This allows the community to exploit peer-to-peer cooperative energy exchanges along with shared neighborhood-level infrastructure to enhance system efficiency.
The strategic energy planning of the community is investigated through the prism of the uncertainty affecting the overall system to ensure a resilient design. 
We thus propose a stochastic problem formulation to identify the optimal design and operation of an energy community accounting for the probability of varying scenarios unfolding.
To ensure the problem remains computationally tractable, we present an uncomplicated distributed optimization scheme, validated by a proof of concept. The distributed sub-problem architecture echoes that of typical decentralized energy management systems, thus anchoring the problem design in a real-world operational control setting, suited for field deployment.
Additionally, this work uncovers the impact of occupant behavior on strategic energy planning thanks to a holistic, context-aware, sensitivity analysis accounting for all key uncertain factors, i.e., climate, economic, and occupants.

In short, the contributions of this work can be summarized as four-fold:
\begin{enumerate}
\item We propose to bridge occupant behavior and strategic urban energy planning by means of an optimal energy community design. Leveraging identified clusters of occupant behaviors along with sub-hourly calibrated building heat dynamics models, we effectively connect granular, detailed spatiotemporal scales of building energy systems to the coarser resolutions commonly employed for urban infrastructure planning.
\item We identify the optimal design and operation of an energy community sensitive to all three system uncertainties, i.e., occupant, climate, and economic, built upon stochastic programming.
\item We propose an instinctive distributed optimization formulation, both securing the computational tractability of the problem, and setting the stage for the decentralized control of the community in a real-world setting.
\item Lastly, we evaluate the impact of occupant behavior on the identified optimal design of the system against other uncertain factors, thus answering the questions: can occupant behavior affect urban energy planning? And, is this effect significant in the context of other system uncertainties?
\end{enumerate}

The remainder of this paper is organized as follows: Sec. \ref{sec:model} presents the energy community system scope and model. In Sec. \ref{sect_methodology} the optimization problem methodology is detailed comprising: the objective (function) as well as the stochastic and distributed problem formulations. Then, implementation specifics are exposed in Sec. \ref{sect_implem} while Sec. \ref{sect_res} reports and discusses the results. Finally, Sec. \ref{sect_con} concludes the article.

    \section{Energy community model}\label{sec:model}
The operating limits of the energy community considered in this study are represented by the energy systems composing the urban energy infrastructure. It comprises groups of individual residential buildings and residents sharing communal resources for the optimal operation and design of the overall community. To appropriately capture the energy dynamics affecting the investigated district, we consider three principal, and connected, modeling blocks; namely, building, grid topology, and community-level system.

\begin{figure}
    \centering
    \begin{adjustbox}{width=0.99\linewidth}
    \includegraphics{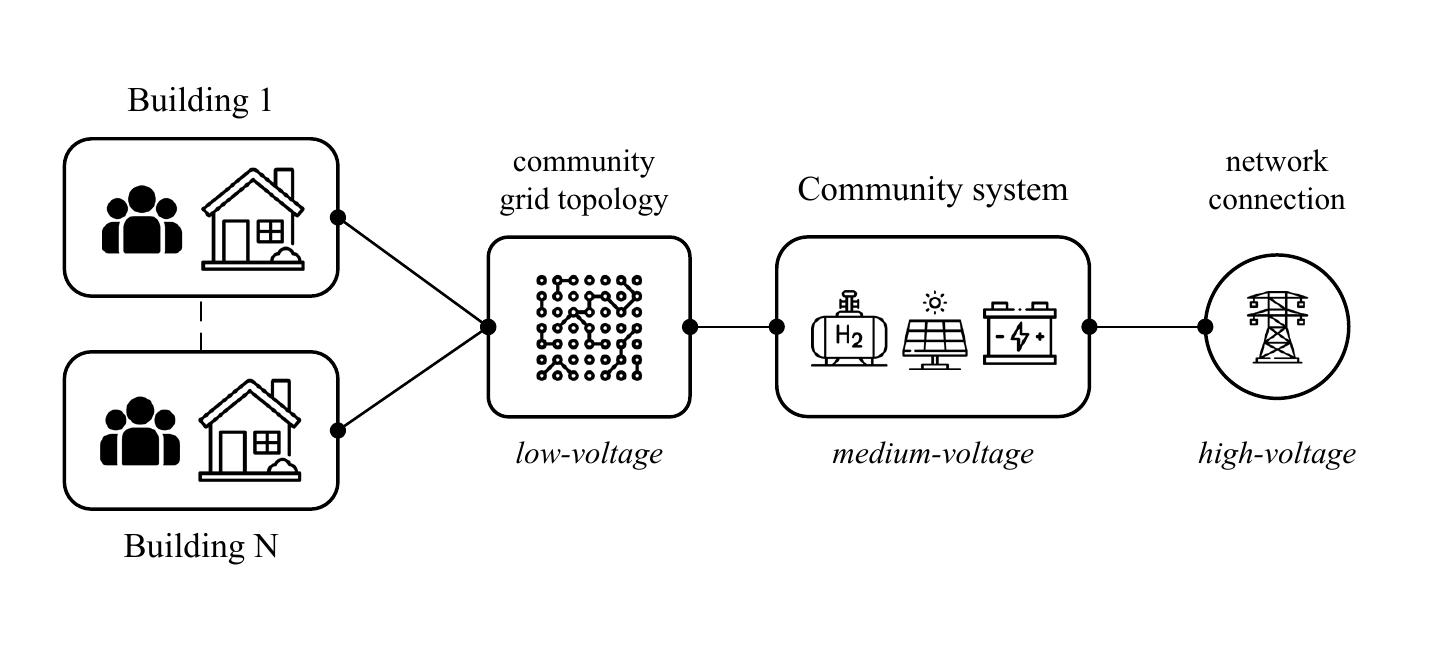}
    \end{adjustbox}
    \caption{Energy community system schematic divided into building, community, and grid topology blocks.}
    \label{fig:model_overview}
\end{figure}

The building block encapsulates residential building utilities providing the electric and heat loads induced by occupant behavior. The models are coupled to detailed thermal characterizations of the building heat dynamics from calibrated stochastic differential equations founded on heat transfer physical laws. This allows the optimization to leverage the thermal inertia of buildings in the operation planning, thus activating their energy flexibility potential.
The grid topology gathers information about the low-voltage electric distribution system connecting the residential buildings together as a community along with power-line and transformer-level power constraints.
Finally, the community system encompasses shared utilities operating on a medium-voltage level while ensuring the overall system connection to the high-voltage distribution grid.
Figure \ref{fig:model_overview} illustrates the energy community modeling blocks schematic.

The behavior of devices and system constraints are modeled using mixed-integer linear programming (MILP), more specifically two-stage stochastic programming. The technique has been widely employed in research to formulate optimization problems as well as perform building services energy optimizations \cite{ashouri2013optimal,schutz2017optimal,fazlollahi2013multi}. 
An advantage of employing MILP stems from the general-purpose solver packages that can be exploited.

\subsection{Formulating conventions}
The following naming convention is used in the rest of the paper:
\begin{itemize}
\item An italic letter stands for a scalar variable, while a bold roman letter represents a vector, commonly indexed over time steps. As an example, the symbols $E$ and $\boldsymbol{E}$ symbolize the electric energy in scalar and vector format, respectively. These annotations are mainly employed to differentiate design variables of devices, i.e., a single value over the entire optimization period, with optimal control equipment variables, i.e., a vector with one value per sample time $t_s$ over the optimization horizon $H$.
\item We differentiate power from energy variables using the time derivative notation $\dot{\boldsymbol{E}}$, here expressing the electric power vector. The relationship between power and energy can be derived using $\boldsymbol{E} =  \dot{\boldsymbol{E}} \cdot t_s$, where $t_s$ is the sampling time. 
\item A superscript $d$ is employed to symbolize independent decision variables. As a result, $\boldsymbol{Q}^d$ indicates a controlled thermal energy manipulated by the optimization.
\item Parameters of the model which depend on uncontrolled variables or external inputs are pre-calculated before the optimization starts. As an example, the COP (coefficient of performance) of the air-source heat pump is a function of ambient temperature, hence it is pre-calculated leveraging weather measurements over a typical meteorological year.
\item The nature of utility investment compels us to employ binary decision variables representing the consideration or disregard of a particular device in the energy community design. For example, the binary decision variable $\chi^d_{\text{U}}$ takes a value of ‘1’ if the unit U is included in the community design and ‘0’ otherwise.
\item All decision variables are declared as non-negative real numbers $\mathbb{R}_{\geq 0}$ such that $\mathbb{R}_{\geq 0} = \{x \in \mathbb{R} \hspace{.1cm}|\hspace{.1cm} x \geq 0\}$.
\end{itemize}

The following sections will describe the structure of the three main model blocks, namely the building, the grid topology, and the community block.

\subsection{Building system}
\begin{figure}
    \centering
    \begin{adjustbox}{width=0.80\linewidth}
    \includegraphics{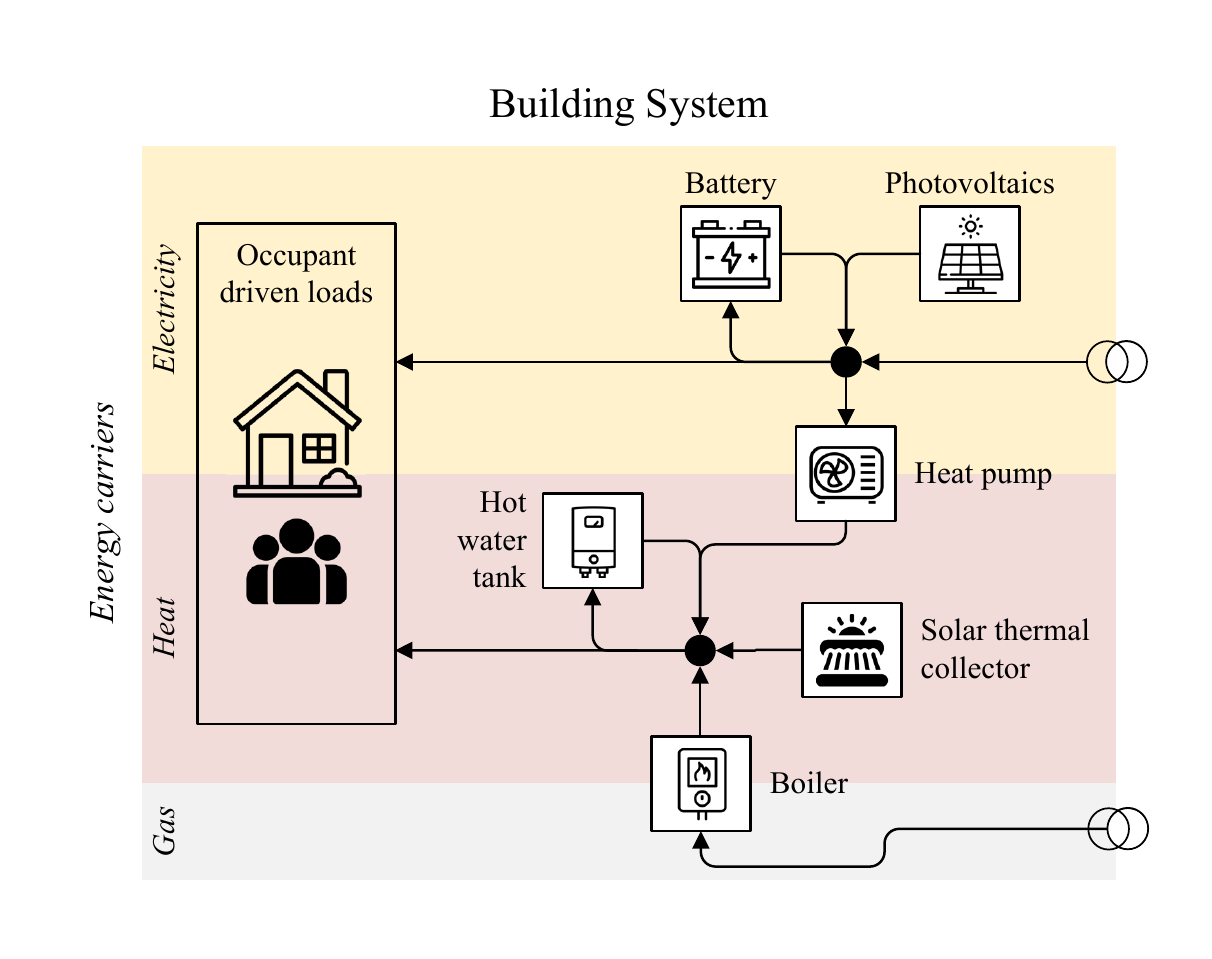}
    \end{adjustbox}
    \caption{Building model block with highlighted energy carriers and connections to the community grid.}
    \label{fig:model_building}
\end{figure}

The building block considers a series of residential storage and conversion technologies commonly employed in Dutch residential homes. Utilities considered englobe solar thermal collectors, photovoltaic panels, storage technologies with a battery and hot water tank, as well as thermal energy converters, i.e., a heat pump, and gas boiler. 
The building thermal dynamics are implemented built upon calibrated lumped resistance capacity models, allowing the controller to leverage the full thermal energy flexibility potential of the communal building stock.
Modeled utilities ensure occupant-driven electric loads and building thermal conditions are met, all the while serving smart community energy management thanks to their connection to the low-voltage distribution network. 
Figure \ref{fig:model_building} illustrates the building model block.

\subsubsection{Heat dynamics model}
Formulating models that support the inclusion of occupant comfort needs while leveraging energy flexibility potential requires an a priori characterization of a building's thermal dynamics. 
This allows the thermal mass of the dwelling to be exploited as a dynamic storage asset.
Conventional building control strategies do not typically consider the thermal mass of the building in their control scheme. Typically, the thermostat is set back to a lower temperature when the building is not occupied such that heating equipments are generally off during these periods. 
However, exploiting the building mass as a thermal storage asset has been shown to significantly reduce operational costs in a context of varying energy prices thanks to load-shifting. These smart control strategies exploit the use of low-cost off-peak electrical energy with improved mechanical heating efficiencies at times where more favorable part-load and ambient conditions occur \cite{braun2003load}.
Additionally, the aggregation of load shifting and load curtailment demand-side management coordinated on a neighborhood scale delivers significant cost reductions \cite{fazeli2011investigating,sheikhi2015integrated}.
Recent study results showed that the implementation of the demand response program significantly reduced the demand for power during peak hours, thereby reducing the installed capacity of the combined heat and power unit \cite{rahgozar2022resilience}.
This highlights the added value brought by considering energy-flexible buildings in the planning phase of building energy systems.

The building thermal models considered in this paper are based on established lumped resistance capacity (RC) models \cite{bacher2011identifying} ranging from 1$^\text{st}$ to 5$^\text{th}$ order. Model parameters were calibrated employing the automated selection and evaluation procedure proposed in \cite{leprince2022fifty}, and later open-sourced \cite{leprince2022grey} providing 225 calibrated models of Dutch residential building heat dynamics.

The building inside temperature state variable $\boldsymbol{T}^d_{b}$, with associated building index $b$ and time step $t$, is determined with
\begin{gather}
    \begin{aligned}
        \boldsymbol{T}^d_{b}(t) =& \boldsymbol{T}^d_{b}(t-1) + \Delta \boldsymbol{T}^d_{b}(\dot{\boldsymbol{Q}}^d_{SP}, \boldsymbol{T}^{amb}, \dot{\boldsymbol{Q}}^{sol}) \text{ ,} 
    \end{aligned} \label{EC_eq1}
\end{gather}
where $\Delta \boldsymbol{T}^d_{b}$ is the incremental heat exchange between the building and ambient conditions defined by the RC model. It is a function of the input space heating decision variable $\dot{\boldsymbol{Q}}^d_{SP}$ and weather conditions, with ambient temperature $\boldsymbol{T}^{amb}$ and solar irradiance $\dot{\boldsymbol{Q}}^{sol}$.
The complete 5$^\text{th}$ order model is formulated as sets of stochastic differential equations describing the building heat flows \cite{leprince2022fifty}, here described in discrete time by
\begin{flalign}
    \Delta \boldsymbol{T}^d_{b} &\equiv \Delta \boldsymbol{T}^i \text{ ,}
\end{flalign}
\begin{flalign}
    \text{Interior: } \Delta \boldsymbol{T}^i &=  \frac{1}{R^{is}C^i}(\boldsymbol{T}^s - \boldsymbol{T}^i)t_s + \frac{1}{R^{im}C^i}(\boldsymbol{T}^m - \boldsymbol{T}^i)t_s \nonumber\\ 
    &+ \frac{1}{R^{ih}C^i}(\boldsymbol{T}^h - \boldsymbol{T}^i)t_s + \frac{1}{R^{ie}C^i}(\boldsymbol{T}^e - \boldsymbol{T}^i)t_s \nonumber\\
    &+ \frac{1}{R^{ia}C^i}(\boldsymbol{T}^{amb} - \boldsymbol{T}^i)t_s + \frac{1}{C^i} A_w \dot{\boldsymbol{Q}}^{sol} t_s \text{ ,} \label{eq:Ti}
\end{flalign}
\begin{flalign}
    \text{Sensor: } \Delta \boldsymbol{T}^s &= \frac{1}{R^{is}C^s}(\boldsymbol{T}^i - \boldsymbol{T}^s)t_s \text{ ,} \label{eq:Ts}
\end{flalign}
\begin{flalign}
    \text{Medium: } \Delta \boldsymbol{T}^m &= \frac{1}{R^{im}C^m}(\boldsymbol{T}^i - \boldsymbol{T}^m)t_s  \text{ ,} \label{eq:Tm}
\end{flalign}
\begin{flalign}
    \text{Heater: } \Delta \boldsymbol{T}^h &= \frac{1}{R^{ih}C^h}(\boldsymbol{T}^i - \boldsymbol{T}^h)t_s + \frac{1}{C^h} \dot{\boldsymbol{Q}}^d_{SP} t_s  \text{ ,} \label{eq:Th}
\end{flalign}
\begin{flalign}
    \text{Envelope: } \Delta \boldsymbol{T}^e &= \frac{1}{R^{ie}C^e}(\boldsymbol{T}^i - \boldsymbol{T}^e)t_s \nonumber\\
    &+ \frac{1}{R^{ea}C^e}(\boldsymbol{T}^{amb} - \boldsymbol{T}^e)t_s + \frac{1}{C^e} A^e \dot{\boldsymbol{Q}}^{sol} t_s \text{ ,} \label{eq:Te}
\end{flalign}
where the subscripts \textit{i}, \textit{s}, \textit{m}, \textit{h}, and \textit{e} point to inside, sensor, medium, heater, and envelope state components respectively.
For a detailed description of the models, the reader is suggested to refer to the work of Bacher and Madsen \cite{bacher2011identifying}.
The temperature is initiated at the set point (Eq.\eqref{eq:Tset}) and is kept within acceptable boundaries to maintain occupant comfort using Eq. \eqref{eq:Tcomf}, 
\begin{flalign}
    \boldsymbol{T}^d_{b}(0) &= \boldsymbol{T}^{set}_{b}(0) \label{eq:Tset} \text{ ,}\\
    \boldsymbol{T}^{set}_{b} - b &\leq \boldsymbol{T}^d_{b} \label{eq:Tcomf} \text{ ,}
\end{flalign}
where $b$ is a buffer parameter commonly set to 0.5 $^\circ$C.

\subsubsection{Battery storage}
Storage devices are modeled employing straightforward state variable update relationships, commonly used in control-oriented frameworks \cite{ashouri2013optimal}. Although more complex and accurate models exist, their consequent additional computational cost should be avoided for large two-stage optimization problems such as urban energy planning.
For instance, the battery $\text{BAT}$ model employed here ignores degradation from charge and discharge cycles as well as synchronous charging and discharging behaviors. 

\begin{flalign}
    \chi^d_{\text{BAT}} \cdot \underline{C}_{\text{BAT}} \leq C^d_{\text{BAT}} \leq \chi^d_{\text{BAT}} \cdot \overline{C}_{\text{BAT}} \label{eq:BAT_maxdesign}
\end{flalign}
\begin{flalign}
    \underline{C}_{\text{BAT}} \leq \boldsymbol{E}_{\text{BAT}} \leq C^d_{\text{BAT}} \label{eq:BAT_design}
\end{flalign}
\begin{flalign}
    \boldsymbol{E}_{\text{BAT}}(t) = \boldsymbol{E}_{\text{BAT}}(t-1) \cdot \sigma_{\text{BAT}} + \dot{\boldsymbol{E}}^d_{BAT,ch}(t) \cdot \eta_{BAT,ch}  \nonumber\\ 
    \hspace{1cm} - \dot{\boldsymbol{E}}^d_{BAT,dch}(t) \cdot \frac{1}{\eta_{BAT,dch}} \label{eq:BAT_sot}
\end{flalign}
\begin{flalign}
    0 \leq \dot{\boldsymbol{E}}^d_{BAT,ch} \leq \gamma_{BAT,ch} \cdot C^d_{\text{BAT}} \label{eq:BAT_design_ch}\\
    0 \leq \dot{\boldsymbol{E}}^d_{BAT,dch} \leq \gamma_{BAT,dch} \cdot C^d_{\text{BAT}} \label{eq:BAT_design_dch}
\end{flalign}
\begin{flalign}
    \boldsymbol{E}_{\text{BAT}}(0) \leq \boldsymbol{E}_{\text{BAT}}(H) \label{eq:BAT_ini}
\end{flalign}

The main design variable is the battery capacity $C^d_{\text{BAT}}$ which sets the limit for the amount of energy stored at any given time in Eq. \eqref{eq:BAT_design}, which is also lower bounded by a minimum energy state-of-charge parameter $\underline{C}_{\text{BAT}}$.
The existence (binary) variable $\chi^d_{\text{BAT}}$ forces the design variable either to zero or within the allowed limits through Eq. \eqref{eq:BAT_maxdesign}.
The battery state-of-charge or stored energy is calculated using Eq. \eqref{eq:BAT_sot}, where $\sigma_{\text{BAT}}$ is the self-discharge rate and $\eta_{BAT,ch}$ and $\eta_{BAT,dch}$ stand for the battery unit charging and discharging efficiencies respectively.
Equations \eqref{eq:BAT_design_ch} and \eqref{eq:BAT_design_dch} restrict the maximum charging and discharging powers using the coefficients $\gamma_{BAT,ch}$ and $\gamma_{BAT,dch}$ respectively, while Eq. \eqref{eq:BAT_ini} proposes a relaxed cyclic constraint for the storage system over the problem horizon $H$.

\subsubsection{Thermal energy storage}
The thermal energy storage (TES), i.e., a hot water tank, is modeled analogously to the battery unit. 
The design variable here is the energy storage capacity $C^d_{\text{TES}}$.

\begin{flalign}
    \chi^d_{\text{TES}} \cdot \underline{C}_{\text{TES}} \leq C^d_{\text{TES}} \leq \chi^d_{\text{TES}} \cdot \overline{C}_{\text{TES}}
\end{flalign}
\begin{flalign}
    0 \leq \boldsymbol{Q}_{\text{TES}} \leq C^d_{\text{TES}}
\end{flalign}
\begin{flalign}
    \boldsymbol{Q}_{\text{TES}}(t) = \boldsymbol{Q}_{\text{TES}}(t-1) \cdot \sigma_{\text{TES}} + \dot{\boldsymbol{Q}}^d_{TES,ch}(t) \cdot \eta_{TES,ch}  \nonumber\\ 
    \hspace{1cm} - \dot{\boldsymbol{Q}}^d_{TES,dch}(t) \cdot \frac{1}{\eta_{TES,dch}}
\end{flalign}
\begin{flalign}
    0 \leq \dot{\boldsymbol{Q}}^d_{TES,ch} \leq \gamma_{TES,ch} \cdot C^d_{\text{TES}} \\
    0 \leq \dot{\boldsymbol{Q}}^d_{TES,dch} \leq \gamma_{TES,dch} \cdot C^d_{\text{TES}}
\end{flalign}
\begin{flalign}
    \boldsymbol{Q}_{\text{TES}}(0) \leq \boldsymbol{Q}_{\text{TES}}(H)
\end{flalign}

\subsubsection{Boiler}
Gas boilers (BOL) are typically employed in Dutch residential heat systems and provide the necessary heat for space heating and hot water demand. Its main design variable is the outlet heating power capacity $C^d_\text{BOL}$.
\begin{flalign}
    \chi^d_\text{BOL} \cdot \underline{C}_\text{BOL} \leq C^d_\text{BOL} \leq \chi^d_\text{BOL} \cdot \overline{C}_\text{BOL}
\end{flalign}
\begin{flalign}
    0 \leq \dot{\boldsymbol{Q}}^d_\text{BOL} \leq C^d_\text{BOL}
\end{flalign}
\begin{flalign}
    \dot{\boldsymbol{Q}}^d_\text{BOL} =  \dot{\boldsymbol{V}}^d_{gas} \cdot \eta_\text{BOL}
\end{flalign}
The output heating power of the boiler $\dot{\boldsymbol{Q}}^d_\text{BOL}$ is obtained by converting input gas $\dot{\boldsymbol{V}}^d_{gas}$ to heat given a fixed unit efficiency $\eta_\text{BOL}$.

\subsubsection{Air source heat pump}
Heat pump technologies have become a popular heating solution for buildings given their high efficiencies and environmental performances \cite{sarbu2014general}. They serve as a sustainable alternative to the gas boiler thanks to reduced operational carbon emissions.
The air source heat pump (HP) is implemented such that

\begin{flalign}
    \chi^d_\text{HP} \cdot \underline{C}_\text{HP} \leq C^d_\text{HP} \leq \chi^d_\text{HP} \cdot \overline{C}_\text{HP} \text{ ,}
\end{flalign}
\begin{flalign}
    0 \leq \dot{\boldsymbol{Q}}^d_\text{HP} \leq C^d_\text{HP}\text{ ,}
\end{flalign}
\begin{flalign}
    \dot{\boldsymbol{Q}}^d_\text{HP} =  \dot{\boldsymbol{E}}^d_\text{HP} \cdot \textbf{COP}_\text{HP} \text{ ,}
\end{flalign}
where $C^d_\text{HP}$ is the design variable and $\dot{\boldsymbol{Q}}^d_\text{HP}$ the output heat power.
The parameter $\textbf{COP}_\text{HP}$ is pre-calculated using an exponential function of the ambient temperature $\boldsymbol{T}^{amb}$ and the, fixed, distribution temperature $T^{dist}$, as defined in Ref. \cite{ashouri2013optimal}:
\begin{flalign*}
    \textbf{COP}_\text{HP} = \alpha_{HP,1} \cdot \text{exp}(\alpha_{HP,2} \cdot (T^{dist} - \boldsymbol{T}^{amb})) \nonumber \\
    + \alpha_{HP,3} \cdot \text{exp}(\alpha_{HP,4} \cdot (T^{dist} - \boldsymbol{T}^{amb})) \text{ ,} \label{eq:HP_COP}
\end{flalign*}
The parameters $\alpha_{HP,*}$ depend on the type of heat pump considered and are provided by the manufacturer.

\subsubsection{Photovoltaic}
Buildings are emerging as growing electricity prosumers who not only produce energy from distributed energy resources but also consume generated energy locally \cite{huang2019transforming}.
With the European Union mandating PV on all commercial, public, and new buildings by 2027 \cite{SaphoraS2022}, photovoltaic systems will soon become an irreplaceable element of our built environment.
We model PV via:
\begin{flalign}
    \chi^d_\text{PV} \cdot \underline{A}_{b} \leq A^d_\text{PV} \leq \chi^d_\text{PV} \cdot \overline{A}_{b} \text{ ,}
\end{flalign}
\begin{flalign}
    \dot{\boldsymbol{E}}_\text{PV} =  A^d_\text{PV} \cdot \boldsymbol{I}^{sol} \cdot \eta_\text{PV} \label{eq:PV_con} \text{ ,}
\end{flalign}
where $A^d_\text{PV}$ is the upper bounded design variable by available building roof surface $\overline{A}_{b}$.
A theoretical limitation for $\overline{A}_{b}$ would be the area of the roof. However, roof obstacles typically result in a few locations becoming unusable for installing PV. Additionally, PV modules are commonly mounted with an inclination, hence its calculations hinge on the geometries of the roof.
The energy conversion equation is straightforwardly implemented in Eq. \eqref{eq:PV_con} employing the nominal efficiency $\eta_\text{PV}$.

\subsubsection{Solar thermal collector}
The solar thermal collector (STC) absorbs sunlight and converts it to heat. The amount of absorbed solar power depends on the collector's surface area $A^d_{STC}$, the total solar incident on the STC surface $\boldsymbol{I}^{sol}$, and the ambient temperature $\boldsymbol{T}^{amb}$, as defined in Ref. \cite{ashouri2013optimal}.

\begin{flalign}
    \chi^d_\text{STC} \cdot \underline{A}_{b} \leq A^d_\text{STC} \leq \chi^d_\text{STC} \cdot \overline{A}_{b}
\end{flalign}
\begin{flalign}
    \dot{\boldsymbol{Q}}_\text{STC} =  A^d_\text{STC} \cdot \eta_\text{STC} \cdot \large( \boldsymbol{I}^{sol} - U_\text{STC} \cdot (T_\text{STC} - \boldsymbol{T}^{amb}) \large) \label{eq:STC}
\end{flalign}
Thermal losses of the collector are modeled in Eq. \eqref{eq:STC} by the term $U_\text{STC} \cdot (T_\text{STC} - \boldsymbol{T}^{amb})$, with $U_\text{STC}$ being the thermal transmittance to the surroundings and $T_\text{STC}$ denoting temperature of the water entering the STC.

Lastly, to consider limited roof area for both PV and STC modules, an upper bound linking both design variables is imposed such that
\begin{flalign}
    A^d_{PV} + A^d_\text{STC} \leq \overline{A}_{b} \text{ .}
\end{flalign}

\subsubsection{Energy balance}
To connect considered devices of the building model block with occupant-driven energy needs a heat (Eq. \eqref{eq:heatbalance}) and electricity (Eq. \eqref{eq:elbalance}) energy balance are modeled as
\begin{flalign}
    \dot{\boldsymbol{Q}}^d_{SP} + \dot{\boldsymbol{Q}}^d_{TES,ch} = \dot{\boldsymbol{Q}}^d_\text{HP} + \dot{\boldsymbol{Q}}^d_\text{BOL} + \dot{\boldsymbol{Q}}^d_{TES,dch}\label{eq:heatbalance} \text{ ,}
\end{flalign}
\begin{flalign}
    \dot{\boldsymbol{E}}^{\text{base}}_{b} + \dot{\boldsymbol{E}}^d_{BAT,ch} + \dot{\boldsymbol{E}}^d_\text{HP} + \dot{\boldsymbol{E}}^d_{b,out} = \dot{\boldsymbol{E}}^d_{BAT,dch} + \dot{\boldsymbol{E}}_{PV} \nonumber \\
    + \dot{\boldsymbol{E}}^d_{b,in}\label{eq:elbalance} \text{ ,}
\end{flalign}
where $\dot{\boldsymbol{E}}^d_{b,in/out}$ stands for the input and output power flows connecting the building model block to the low-voltage grid.
The left-hand side elements of both equations denote the energy demands of the building and its utilities while the right-hand side elements provide the required energy to meet the demands. 
It can here be noted that while the building's space heat load $\dot{\boldsymbol{Q}}^d_{SP}$ is optimally controlled by the optimization, as a result of ensuring suitable thermal condition (Eq. \eqref{eq:Tcomf}), the baseline electricity load $\dot{\boldsymbol{E}}^{\text{base}}_{b}$ associated with occupant-behavior is a fixed, non-shiftable, load measurements.

\subsection{Community system}
\begin{figure}
    \centering
    \begin{adjustbox}{width=0.70\linewidth}
    \includegraphics{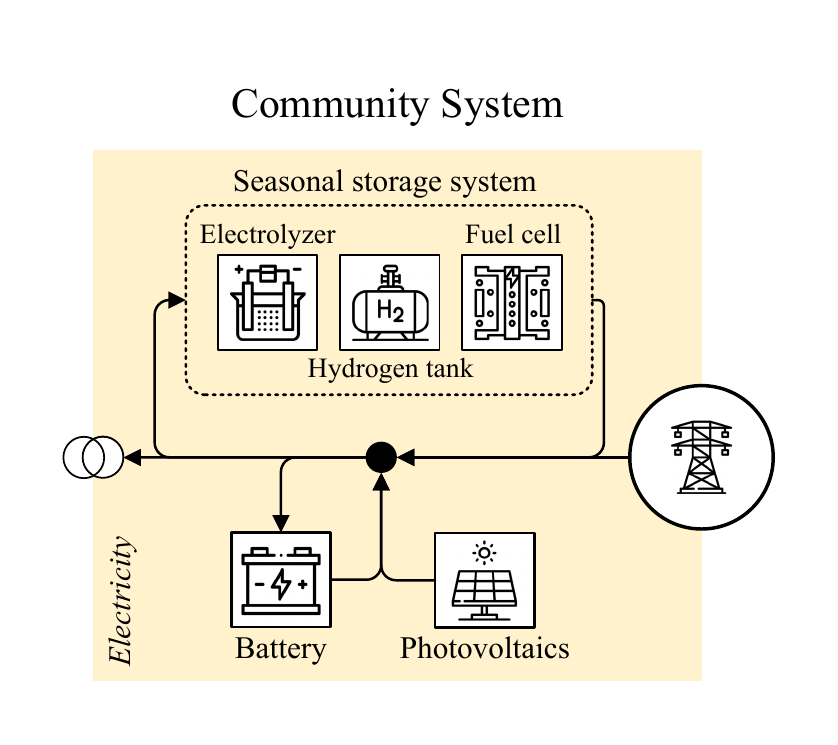}
    \end{adjustbox}
    \caption{Community model block connected to the low-voltage distribution network towards buildings (left-hand side) and to the high-voltage distribution grid (right-hand side).}
    \label{fig:model_community}
\end{figure}

The community system exemplifies the concept of the energy hub, operating at a medium voltage network scale, linking the building community to a shared set of utilities with the high-voltage energy grid (Fig. \ref{fig:model_community}). In this setting, the community system considers utilities that might benefit from increased performances due to their larger capacities, namely photovoltaics coupled with short and/or seasonal storage systems.
The seasonal storage system is composed of three devices set up in series, i.e., an electrolyzer converting electricity to hydrogen, a hydrogen tank for long-term energy storage, and a fuel tank converting hydrogen back to electricity \cite{murray2018comparison}. 

Models of the photovoltaic and battery community system devices are analogous to the ones presented in the building block. The main differentiation between them stems from their techno-economic parameters. 
We detail the particularities of the seasonal storage device to explicitly illustrate the sizing of three separate entities under a unified storage utility.

\subsubsection{Seasonal storage system}
The important value brought by the consideration of seasonal storage devices originates from offsetting seasonal mismatches between renewable energy generation and energy demands. With hydrogen storage tanks featuring negligible energy losses, they are popularly considered a promising solution for long, inter-seasonal, storage systems \cite{gabrielli2018optimal}.
The hydrogen tank (HYD) is coupled to the electrolyzer (EL) and fuel cell (FC) to produce, store, and use hydrogen respectively.
A compressor device is connected to the storage tank to store hydrogen at a high pressure of 200 bars, and while hydrogen storage possesses limited energy losses, the round-trip efficiency of the seasonal storage system is much lower than that of the battery, i.e., about 35\% against 95\% respectively.
For these reasons, hydrogen storage has been investigated as an efficient alternative to store energy for long periods of time \cite{gabrielli2018optimal,darivianakis2017data}.
It is finally worth mentioning that due to the differences in usage between batteries and seasonal storage systems, this typically translates into larger installed capacities for seasonal storage devices.

The seasonal storage system is modeled analogously to other storage devices with the addition of three distinct design variables $C^d_{\text{HYD}}$, $C^d_{\text{EL}}$, and $C^d_{\text{FC}}$ standing for the hydrogen tank, electrolyzer and fuel cell respectively, all linked by a unique existence variable $\chi^d_{\text{HYD}}$ in Eqs. \eqref{eq:HYD_maxdesign1}, \eqref{eq:HYD_maxdesign2}, and \eqref{eq:HYD_maxdesign3}.
\begin{flalign}
    \chi^d_{\text{HYD}} \cdot \underline{C}_{\text{HYD}} \leq C^d_{\text{HYD}} \leq \chi^d_{\text{HYD}} \cdot \overline{C}_{\text{HYD}} \label{eq:HYD_maxdesign1}\\
    \chi^d_{\text{HYD}} \cdot \underline{C}_{\text{EL}} \leq C^d_{\text{EL}} \leq \chi^d_{\text{HYD}} \cdot \overline{C}_{\text{EL}} \label{eq:HYD_maxdesign2}\\
    \chi^d_{\text{HYD}} \cdot \underline{C}_{\text{FC}} \leq C^d_{\text{FC}} \leq \chi^d_{\text{HYD}} \cdot \overline{C}_{\text{FC}} \label{eq:HYD_maxdesign3}
\end{flalign}
\begin{flalign}
    \underline{C}_{\text{HYD}} \leq \boldsymbol{E}_{\text{HYD}} \leq C^d_{\text{HYD}} \label{eq:BAT_design}
\end{flalign}
\begin{flalign}
    \boldsymbol{E}_{\text{HYD}}(t) = \boldsymbol{E}_{\text{HYD}}(t-1) \cdot \sigma_{\text{HYD}} + \dot{\boldsymbol{E}}^d_{EL,ch}(t) \cdot \eta_{EL,ch}  \nonumber\\ 
    \hspace{1cm} - \dot{\boldsymbol{E}}^d_{FC,dch}(t) \cdot \frac{1}{\eta_{FC,dch}} \label{eq:HYD_sot}
\end{flalign}
\begin{flalign}
    0 \leq \dot{\boldsymbol{E}}^d_{EL,ch} \leq \gamma_{EL,ch} \cdot C^d_{\text{EL}} \label{eq:HYD_design_ch}\\
    0 \leq \dot{\boldsymbol{E}}^d_{FC,dch} \leq \gamma_{FC,dch} \cdot C^d_{\text{FC}} \label{eq:HYD_design_dch}
\end{flalign}
\begin{flalign}
    \boldsymbol{E}_{\text{HYD}}(0) \leq \boldsymbol{E}_{\text{HYD}}(H)
\end{flalign}
The electrolizer and fuel cell fix the charging $\eta_{EL,ch}$ and discharging $\eta_{FC,dch}$ efficiencies of the storage tank, and limit its inlet $\dot{\boldsymbol{E}}^d_{EL,ch}$ and outlet $\dot{\boldsymbol{E}}^d_{FC,dch}$ powers through Eqs. \eqref{eq:HYD_design_ch} and \eqref{eq:HYD_design_dch} respectively.

\subsubsection{Power balance}
The power balance equation linking the community system devices together with the low- and high-voltage distribution energy grids is modeled as
\begin{flalign}
      \dot{\boldsymbol{E}}^d_{MV \rightarrow LV} + \dot{\boldsymbol{E}}^d_{BAT,ch} + \dot{\boldsymbol{E}}^d_{EL,ch} = \dot{\boldsymbol{E}}^d_{LV \rightarrow MV} + \dot{\boldsymbol{E}}_{PV,in} + \dot{\boldsymbol{E}}^d_{BAT,dch} \nonumber \\
      + \dot{\boldsymbol{E}}^d_{FC,dch} + \dot{\boldsymbol{E}}^d_{HV,in} \text{ ,}
\end{flalign}
where HV represents the input high-voltage power flow and $MV \rightarrow LV$ and $LV \rightarrow MV$ stand for the medium-to-low and low-to-medium voltage network connections respectively. BAT, PV, EL, and FC refer to battery storage, photovoltaics, electrolyzer, and fuel cell utilities on a community level, respectively.

\subsection{Grid topology}
The topology of the low-voltage distribution network, connecting the buildings forming the energy community together is here presented. 
\begin{flalign}
      \dot{\boldsymbol{E}}^d_{MV \rightarrow LV/LV \rightarrow MV} &\leq \overline{\dot{E}}_{\text{MV}} + s^d_{MV} \label{eq:max_MV} \\
       \dot{\boldsymbol{E}}^d_{b,in/out} &\leq \overline{\dot{E}}_{\text{LV}} + s^d_{LV,b} \hspace{1cm} \forall b \in \mathbb{B} \label{eq:max_LV}
\end{flalign}
\begin{flalign}
      \dot{\boldsymbol{E}}^d_{MV \rightarrow LV} + \sum_{\text{b} \in \mathbb{B}} \dot{\boldsymbol{E}}^d_{\text{b},out} = \sum_{\text{b} \in \mathbb{B}} \dot{\boldsymbol{E}}^d_{\text{b},in} + \dot{\boldsymbol{E}}^d_{LV \rightarrow MV} \label{eq:COM_balance}
\end{flalign}
In a typical distribution network, power flows are limited by one of two factors: the maximum capacity of the power lines (Eq. \eqref{eq:max_MV}) or the maximum capacity of the micro-grid transformer (Eq. \eqref{eq:max_LV}), here represented by $\overline{\dot{E}}_{\text{MV}}$ and $\overline{\dot{E}}_{\text{LV}}$ respectively.
Penalized slack variables $s^d$ are additionally included to relax both LV and HV line maximum capacities in order to secure problem feasibility.
Each individual building of the energy community belongs to the set $\mathbb{B}$ such that Eq. \eqref{eq:max_LV} holds for all $b \in \mathbb{B}$ and the power balance of Eq. \eqref{eq:COM_balance} sums the in and output power of all the buildings belonging to the community.

    
    \section{Methodology}\label{sect_methodology}
The principal objective of the energy community optimization problem is to identify the optimal selection, sizing, and operation of available building and community-level components while accounting for the uncertainty affecting the system in order to minimize the total communal costs.
Optimal system designs are, however, inherently co-dependent on their associated operational strategy.
As such, simultaneous optimization of control and design approaches \cite{bansal2002case} perform well for one realization of the uncertainty affecting the system, but may not for others. These approaches imply that if the suggested design is used, no other control strategy yields better results, i.e., a lower cost function, and vice versa.
Designing an energy system that is \text{resilient} to these uncertainties is an important goal of this work. 

This section consequently presents the objective function of the optimization problem and details how the uncertainty affecting the energy community is captured into representative scenarios. 
Then, uncertainty is incorporated into the optimization problem as a two-stage stochastic programming model, and a sensitivity analysis is proposed to evaluate the specific contribution of varying uncertainty factors on the optimal energy community system design, in particular the occupant behavior. 
Lastly, a distributed formulation of the problem is put forth dealing with computational tractability issues endowed from large and granular optimization problems.

\subsection{Objective function}
Considering the optimal energy planning goal of the considered community, the optimization is performed for a full year encompassing all seasonal variations, while the objective is extended to varying equipment lifetimes ranging from 10 to 25 years. 
The total objective function $O^{tot}$ to be minimized consists of three terms associated with levelized investment, operation, carbon emission reduction objectives, and slack penalties:
\begin{flalign}
    \text{min} \hspace{1cm} O^{tot} = O^{inv}_{lvl} + O^{opr} + O^{co2} + O^{slk} \text{ .}
\end{flalign}

The levelized investment objective $O^{inv}_{lvl}$ is calculated in Eq. \eqref{eq:O_inv} as the sum of all levelized price variables $p_{U}$, which indicates the overall cost of purchase, installation, maintenance and replacement of an arbitrary unit $U$ belonging to the building and community system sets $\mathbb{B}$ and $\mathbb{C}$ respectively.
The prices are levelized over the technology lifetime $\tau_{U}$ at a discount rate $r$ such that their operation horizons serve as weights for their investment costs in the optimization.

\begin{flalign}
    & O^{inv}_{lvl} = \sum_{U \in \mathbb{B} \cup \mathbb{C} } p_{U} \cdot \frac{r}{1-(1+r)^{-\tau_{U}}} \label{eq:O_inv}
\end{flalign}
\begin{flalign}
    p_{U} = a_{U} \cdot D^d_{U} + b_{U} \cdot \chi^d_{U} \label{eq:p_U}
\end{flalign}
The device price $p_{U}$ is affected by the existence variable $\chi^d_{U}$ and the main design variable $D^d_{U}$ of the device, which refers either to the capacity $C^d_{U}$ or area $A^d_{U}$ of the unit. Equation \eqref{eq:p_U} shows the calculation of the price for including a device U in the energy community.
The parameter $b_{U}$ defines the price for the existence of the device, while the parameter $a_{U}$ represents the relative sizing price of the unit.

The operational cost $ O^{opr}$ accounts for the total amount of electricity and gas consumed by the energy community.
\begin{flalign}
    & O^{opr} = \dot{\boldsymbol{E}}^d_{\text{HV}} \cdot \boldsymbol{p}^T_{el} + \sum_{b \in \mathbb{B}} \dot{\boldsymbol{V}}^d_{gas,b} \cdot \boldsymbol{p}^T_{gas} \label{eq:O_opr}
\end{flalign}

The carbon emission reduction objective $O^{co2}$ is modeled as a carbon emission penalty associated with the natural gas consumption of buildings. The term $\boldsymbol{p}_{co2}$ indicates the carbon pricing set by the European Union. 
\begin{flalign}
    & O^{co2} = \sum_{b \in \mathbb{B}} \dot{\boldsymbol{V}}^d_{gas,b} \cdot \boldsymbol{p}^T_{co2} \label{eq:O_co2}
\end{flalign}

Lastly, the slack penalty costs $O^{slk}$ associate a predefined slack penalty $p_{slk}$ to all declared slack variables, such that
\begin{flalign}
    O^{slk} = p_{slk} \cdot s^d_{MV} + p_{slk} \cdot \sum_{b \in \mathbb{B}} s^d_{LV,b} \text{ .} \label{eq:O_slk}
\end{flalign}
The penalty value is set high such that the optimization would only consider relaxing the constraints in cases of problem infeasibility. 

It should be noted that the energy community optimization problem shares a unique, global objective function $O^{tot}$. 
Defining a global objective function ensures a cooperative behavior between all elements of the community, i.e., building and energy community blocks, working toward the reduction of the aggregated costs of the system, rather than sub-optimal individualistic objectives. Leveraging such cooperative behaviors between individual agents of an energy system is recognized to substantially improve economic and energetic performances \cite{van2020cooperative}.

\subsection{Representative scenario identification}
To perform urban energy planning in a computationally tractable manner, it becomes necessary to trim encompassed spatiotemporal dimensions to a reduced, but representative, number of spatiotemporal frames.
Indeed, urban distributed energy resources design procedures commonly cluster encompassed input data to a typical reference year, assumed constant over the lifetime of the energy system, e.g., 25 years \cite{rager2015urban}.
Downscaling the spatial resolution of the energy community by clustering buildings to fewer representative ones would, however, deprive the optimization of the diversity of information-rich occupant behaviors and varying building thermal dynamics.
The purpose of the present work is to consider the complete building community stock in the energy planning process, allowing building energy flexibility activations to be exploited on an aggregated urban scale while bridging the pluralities of occupants to urban infrastructure planning.
The determination of characteristic years of measurements across buildings, weather, and economic conditions provides scenarios serving both two-stage stochastic programming (Sec. \ref{sec:stochastic-program}) and the latter sensitivity analysis (Sec. \ref{sec:sensitivity-analysis}).
In order for these scenarios to approximate the underlying uncertainty as closely as possible, many scenarios are initially bootstrapped (Sec. \ref{sec:scengen}), then reduced to few representative ones by clustering (Sec. \ref{sec:scenred}).

\subsubsection{Scenario generation using seasonal bootstrapping} \label{sec:scengen}
In this study, three categories of parameters are selected as uncertain, namely building electrical load demands and set-point temperatures for occupant behavior, energy prices for economic conditions, and ambient temperature and solar irradiance for weather conditions. 

To artificially increase the number of years of collected data while retaining the auto-correlations of energy consumption profiles and day-ahead pricing, we apply a seasonal block bootstrapping technique to generate 1000 synthetic years of data.
Block bootstrapping for seasonal time series has been found suitable for periodic time series with fixed-length periodicities of arbitrary block and sample size \cite{dudek2014generalized}.
Given the diurnal patterns of building energy consumption and day-head electric forecasts, we consider block samples of 24 hours that are sampled across the entire data set to secure the correlations between building energy needs and weather and economic conditions.
The blocks are bootstrapped over a seasonal-dependent sub-space to retain the periodic behaviors present in the original data. Thereby, weekday and weekend variations are preserved while the sampling space is restricted to a region of 8 weeks surrounding the sampling block \cite{https://doi.org/10.48550/arxiv.2211.00934}.

\subsubsection{Scenario reduction using clustering} \label{sec:scenred}
Gathered scenarios are then reduced to a more manageable number employing k-medoids clustering \cite{park2009simple} to obtain identifiable cluster centers (\textit{medoids}) and associated probabilities \cite{en11123310}.
The advantage of uncovering medoids, which are superimposed on existing input data, is that it preserves the volatility of the original input data as opposed to k-means clustering which produces \textit{centroids} that are averages of their cluster members, thus resulting in the curtailment of their individual stochastic properties.

Identifying a suitable number of clusters is commonly performed from cluster intra-class homogeneity and inter-class separation indexes that assess the validity of obtained clusters.
These metrics, however, deliver no information on the grouping validity of the resulting optimal policy of an optimization problem. 
This is why any arbitrary number of clusters can be considered a suitable identification of representative scenarios, given a sufficient number of samples.
We consequently reduce the 1000 bootstrapped scenarios to $N_\Omega=10$ distinct clusters, where $N_\Omega$ is the number of scenarios considered and $\Omega$ is the set of scenarios.
Their associated probabilities $\pi(w)$ is subsequently obtained from
\begin{flalign}
    \pi(\omega) = P(\omega | \lambda = \lambda(\omega)), \hspace{.3cm} \text{where} \hspace{.3cm} \sum_{\omega \in \Omega} \pi(\omega) = 1 \text{ ,}
\end{flalign}
and $\lambda(\omega)$ is a random variable associated with a scenario index $\omega$, while the scenario realization probabilities are represented by $\pi(\omega)$. The probabilities $\pi(\omega)$ correspond to the relative cluster sizes obtained via k-medoids clustering.

\subsection{Stochastic programming formulation} \label{stochastic-program}
Introducing uncertainty in the design of energy communities involves a decision-making problem structure well suited to a two-stage stochastic programming model \cite{mavromatidis2018design}.
Indeed, the design problem features the concurrent determination of both design and operation variables, which are commonly decided in different stages.
This means that decisions on the design variables must be adequate to adapt to varying realizations of energy demand and supply profiles over the year.
Consequently, design variables $D_d$ and their associated existence variables $\chi^d$ are categorized as first-stage variables, to be decided prior to the resolution of uncertainty, and all other operational (decision) variables are considered second-stage variables, which can later be adapted in function of the uncertainty scenario unfolding.
A two-stage stochastic programming approach thus undertakes the simultaneous determination of the optimal configuration of a distributed energy system given varying (optimal) operating conditions \cite{zhou2013two}.

By incorporating the uncertainty into the mixed-integer optimization problem, a two-stage stochastic programming problem is formulated as follows \cite{conejo2010decision}:
\begin{gather}
    \begin{aligned}
        \text{min} & \overbrace{O^{inv}_{lvl}}^{\text{1\textsuperscript{st}-stage costs}} + \overbrace{\sum_{\omega \in \Omega} \pi(\omega) \cdot (O^{opr}(\omega) + O^{co2}(\omega) + O^{slk}(\omega))}^{\text{expected 2\textsuperscript{nd}-stage costs}} \text{ ,}  \\
        s.t. & \hspace{.2cm} \boldsymbol{A}\boldsymbol{x}^d = \boldsymbol{b} \text{ ,}\\
        & \hspace{.2cm} \boldsymbol{T}(\omega)\boldsymbol{x}^d + \boldsymbol{W}(\omega)\boldsymbol{y}^d(\omega) = \boldsymbol{h}(\omega) \hspace{.2cm} \forall \omega \in \Omega \text{ ,}
    \end{aligned}\label{stochastic_problem}
\end{gather}
where $\boldsymbol{x}^d$ gathers the 1\textsuperscript{st}-stage decision variables and $\boldsymbol{y}^d(\omega)$ the 2\textsuperscript{nd}-stage decisions. 
The matrices and vectors $\boldsymbol{A}$, $\boldsymbol{T}(\omega)$, $\boldsymbol{W}(\omega)$, $\boldsymbol{b}$, $\boldsymbol{c}$, $\boldsymbol{q}(\omega)$, and $\boldsymbol{h}(\omega)$ are known parameters of the system, that can be gathered from Eqs. \eqref{EC_eq1}-\eqref{eq:COM_balance} and \eqref{eq:O_inv}-\eqref{eq:O_co2}.

The objective thus becomes to determine the first-stage design and existence variables by taking the sum of the deterministic first-stage costs, namely, the levelized investments costs, defined in Eq. \eqref{eq:O_inv}, and the expected second-stage operational costs corresponding to the sum of the operation and carbon emission costs, see Eqs. \eqref{eq:O_opr} and \eqref{eq:O_co2} respectively, weighted by their respective realization probabilities $\pi(\omega)$.

\subsection{Uncertainty impact on energy community design} \label{sensitivity-analysis}
While the consideration of the uncertainty in the form of a stochastic program allows the identification of the optimal policy given the probability of varying scenarios unfolding, it, however, does not inform energy planners on the relative \textit{impact} of its considered uncertainty factors.
To evaluate the relative influence of the different categories of uncertain parameters on the design of energy communities, in particular occupant behavior, it becomes necessary to undertake a sensitivity analysis.
There are two principal methods for sensitivity analysis, local and global ones.
Local sensitivity analysis methods typically analyze how the uncertainty in each input parameter affects an output of interest \cite{morio2011global}. Uncertain parameters are commonly altered one at a time with other parameters fixed at their nominal values, or through the definition of scenarios, i.e., combinations of uncertain parameter values \cite{moret2017strategic}.
Global sensitivity analysis on the other hand considers all of the input parameters simultaneously.
The impact of each input parameter on the performance indicator of interest are commonly evaluated by variance-based methods. These are, however, computationally expensive for urban energy planning due to their large number of inputs \cite{kucherenko2009monte}, and require the characterization of uncertainties a priori, else would result in false rankings when employing generic uncertainty ranges \cite{moret2017characterization}.

To keep the scope of this work within manageable limits, we consider a local sensitivity analysis method performed over all-encompassed uncertainty factors.
This supports the assessment of the impact of occupant behavior on urban energy planning while providing a \text{relative} evaluation in the context of other uncertainties.

\subsubsection{Local sensitivity analysis}
To undertake the sensitivity analysis, uncertainty factor-dependent scenario sub-sets and their nominal values must first be identified.
The scenario ensemble is thus divided into three distinct subsets
\begin{flalign}
    \Omega = \Omega^{occ} \cup \Omega^{eco} \cup \Omega^{clim} \text{ ,}
\end{flalign}
corresponding to occupant, economic, and climate conditions respectively.
The nominal scenario $\omega_{nom}$ for each uncertainty factor is identified from k-medoid clustering with $k=1$ subsets of $\Omega$ and extracting its medoid scenario.
This is performed over the subsets $\Omega^{eco} \cup \Omega^{clim}$, $\Omega^{occ} \cup \Omega^{clim}$, and $\Omega^{occ} \cup \Omega^{eco}$ for occupant, economic, and climate conditions uncertainty factors respectively.

Then, the influence of varying uncertainty factor-dependent scenarios is assessed in a one-at-a-time fashion via the optimal design variables retained by the energy community planning problem.
The problem \eqref{stochastic_problem} is then iteratively solved for either of the following variations
\begin{flalign}
    \Omega_{ijl} = \Omega^{occ}_{i} \cup \Omega^{eco}_{j} \cup \Omega^{clim}_{l} \hspace{.1cm}
    \begin{cases}
        \forall i \in \Omega^{occ}, j = \omega_{nom}^{eco}, l = \omega_{nom}^{clim} \text{ ,}\\
        \forall j \in \Omega^{eco}, i = \omega_{nom}^{occ}, l = \omega_{nom}^{clim} \text{ ,}\\
        \forall l \in \Omega^{clim}, i = \omega_{nom}^{occ}, j = \omega_{nom}^{eco} \text{ .}
    \end{cases}
\end{flalign}
Note that the evaluated set $\Omega_{ijl}$ becomes singular, thus the stochastic program \eqref{stochastic_problem} becomes a deterministic problem as the 1\textsuperscript{st} and 2\textsuperscript{nd} stage costs are evaluated over a unique scenario.

It should be acknowledged, however, that such a setup disregards the existing inter-correlations between the considered uncertain parameters.
For instance, the energy flexibility leveraged in demand-side management applications from occupant-established comfort buffer regions is known to possess a strongly correlated relationship to the energy price levels \cite{AMARALLOPES2020115587}. Similarly, weather conditions typically affect occupant thermal preferences.
This implies that separating these uncertainties in factor-dependent scenarios is intrinsically flawed, making it tedious to differentiate their independent contributions to the problem policy.
We consider this approximation, however, to be a necessary simplification for the evaluation of the distinct impact of these uncertainties.

\subsection{Distributed optimization}
Stochastic optimization problems are notoriously known for their associated computational burden \cite{marin2020novel,toubeau2017medium,zheng2014stochastic}. As problems get larger, the state space is proportionally multiplied by the number of considered scenarios, and computing intractability problems quickly arise.
To alleviate this charge, we propose to partition the problem into smaller sub-problems. 
This allows scaling of the considered energy community system as sub-problems are easier to solve. simultaneously, this increases the overall system resilience in a control setting should one of its components (sub-problems) fail or become obsolete.
The computational load of the problem is subsequently eased by dividing the initially larger problem into multiple smaller ones \cite{costanzo2013coordination}, resulting in a \textit{distributed} optimization problem.

The subsequent partitioned problems, however, require careful coordination not to result in conflicting local actions and threaten the global system stability.
To this end, we consider an uncomplicated sequential solving approach \cite{stadler2016distributed}, allowing information exchange between the different sub-systems while keeping the complexity inherent to each model undisclosed \cite{stephant2021distributed}.
This results in an information-optimized communication system requiring the sole transfer of aggregated local energy flows between sub-systems in the coupling constraint.
This setup notably preserves the privacy related to individual building energy demands, while preparing the stage for the real-world deployment of the energy management system of the community.

The stochastic optimization problem is thus divided into sub-systems to form a distributed stochastic optimization problem. We consider individual building systems coupled with community-level utilities as sub-problems in order to provide the solver with available information from all spatial scales of the system. The grid topology energy balance,  Eq. \eqref{eq:COM_balance}, here serves as an evident coupling constraint, and becomes
\begin{flalign}
      \dot{\boldsymbol{E}}^d_{MV,out} + \sum_{\text{b} \in \mathbb{B} \setminus \{\text{BLG}\} } \dot{\boldsymbol{E}}_{\text{b},out} + \dot{\boldsymbol{E}}^d_{\text{BLG},out} = \nonumber \\ 
      \sum_{\text{b} \in \mathbb{B} \setminus \{\text{BLG}\}} \dot{\boldsymbol{E}}_{\text{b},in} + \dot{\boldsymbol{E}}^d_{MV,in} + \dot{\boldsymbol{E}}^d_{\text{BLG},in} \text{ ,}
\end{flalign}
where BLG is the considered building sub-system being optimized, and $b$ indicates all other buildings belonging to the building set $\mathbb{B}$. Notice how the aggregated energy demands of other building systems $\sum_{\text{b} \in \mathbb{B} \setminus \{\text{BLG}\} } \dot{\boldsymbol{E}}_{\text{b}}$ is now a parameter of the optimization problem, rather than a decision variable.

\begin{figure}
    \centering
    \begin{adjustbox}{width=0.80\linewidth}
    \includegraphics{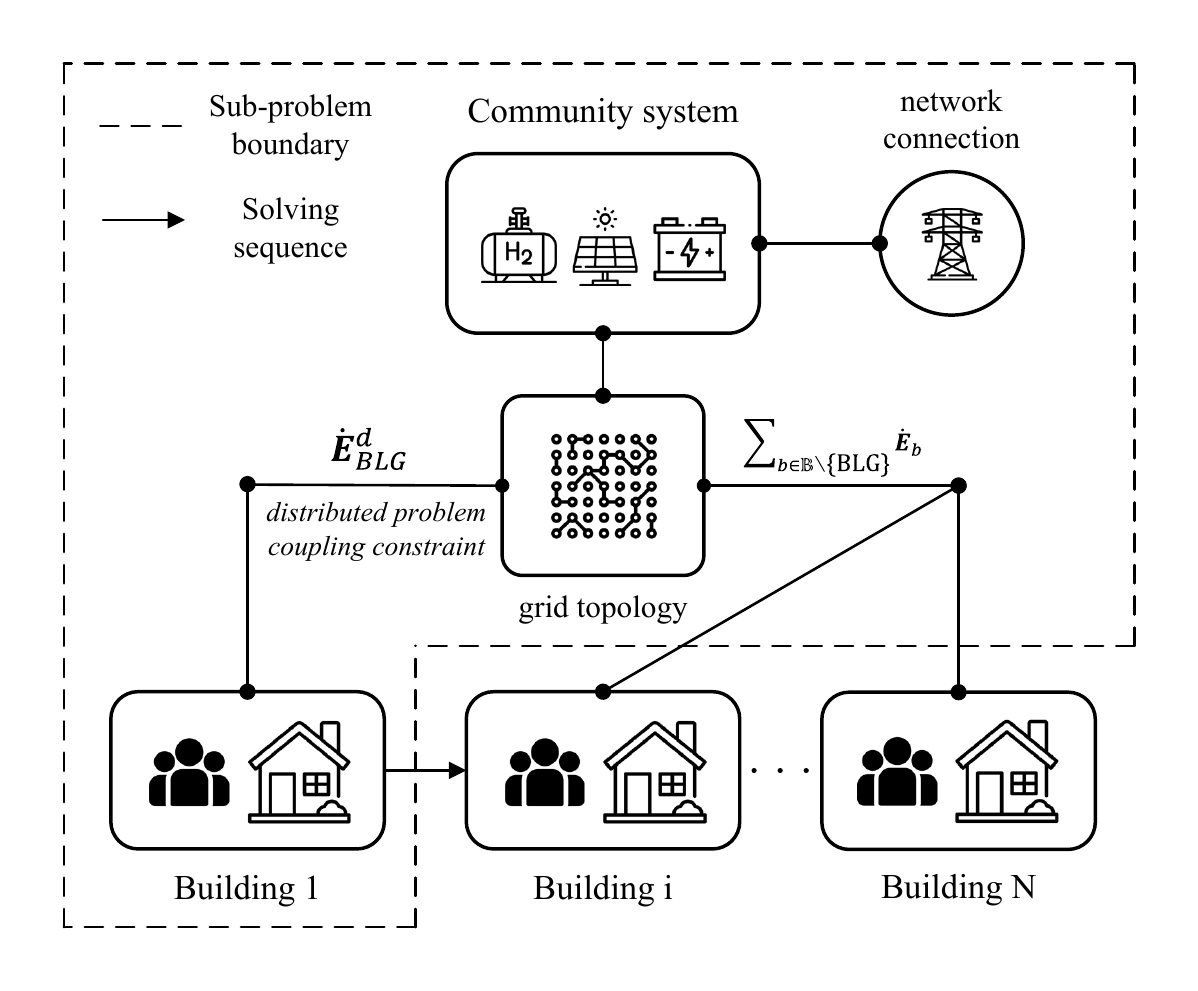}
    \end{adjustbox}
    \caption{Sequential solving scheme of the distributed stochastic optimization problem.}
    \label{fig:distributed_problem}
\end{figure}

The distributed optimization setup is illustrated in Fig. \ref{fig:distributed_problem}.
The problem is iteratively solved until variations of the global objective function are below a predefined threshold $\varepsilon$ such that $\Delta O^{tot} \leq \varepsilon$ defines the stopping criterion of the distributed optimization.

Although the proposed distributed setup lacks a formal mathematical decomposition that would secure the convergence of the problem to the global optimal, we instead undertake a proof of concept, which compares a reduced problem of the proposed distributed stochastic optimization with its centralized counterpart.
Table \ref{tab:poc} summarizes its result and system parameters.
\begin{center}
	\begin{table}
	\caption{Distributed problem proof of concept {\label{tab:poc}}}
	\vspace{-0.5em}
	\setlength\extrarowheight{-3pt}
	\begin{adjustbox}{width=0.96\linewidth}
	    \begin{tabular}{r c c c c }
\toprule
  Problem type & \multicolumn{3}{c}{Parameters} & Objective value \\
  & buildings [$\#$] & scenarios [$\#$] & iterations [$\#$] & [EUR] \\
\midrule
 centralized & 5 & 10 & 1  & 15'188.173 \\
 distributed & 5 & 10 & 19 & 15'188.173 \\
\bottomrule
\end{tabular}
	\end{adjustbox}
	\end{table}
\end{center}
The proof of concept demonstrates that the distributed setup converges to the global optimal solution ensuing the first iteration as a result of the simple, individualistic optimal strategy identified.
Typically, energy exchanges between the varying sub-systems of the distributed problem would iteratively converge to the global optimal within a 1\% margin.
We consider such optimal-close solutions satisfactory, and in fact valuable to the scientific and research community as these provide a simple and intuitive problem distribution arrangement supporting scalable strategic urban energy planning, thus facilitating the accessibility of our approach.
Additionally, the structure of the distributed problem may be subsequently employed for the decentralized control of the energy community, by simply disregarding the investment-related variables.

\section{Implementation}\label{sect_implem}
The energy community problem is implemented in Python using the PuLP package \cite{mitchell2011pulp} as an interface to the Gurobi solver \cite{gurobi}. 

For the energy community system considered, we employ real historical building measurements collected from the energy distributor Eneco, thus anchoring our approach on data-driven techniques to induce realistic results.
A total of 225 homes located in the Netherlands, a European region under the K\"oppen climate classification index \cite{CHEN201369} \textit{Cfb} which describes mild temperate, fully humid, and warm summer regions, are treated.
Anonymized measurements are gathered from smart thermostats temperature set-points and electrical meter consumption collected by the energy distributor Eneco at resolutions of 10 seconds and 15 minutes respectively, over a period of 3 years starting from January 1\textsuperscript{st} 2019 to the 2\textsuperscript{nd} of December 2022.
Their associated building heat dynamics models are extracted from the open data set Grey-Brick Buildings \cite{leprince2022grey} established from the same case study. 
We filter out models exhibiting nCPBES (normalized cumulated periodogram boundary excess sum) higher than 0.01 to retain models of good fit quality exclusively, resulting in 41 remaining buildings.
While these buildings are located in varying regions of the Netherlands, we propose to construct a synthetic neighborhood by unifying their collected information under a common atmospheric condition. Given the homogeneity of the Dutch geographical climate, we consider this assumption to be a suitable approximation.

Weather data is assembled from publicly available Royal Netherlands Meteorological Institute (KNMI) weather station measurements \cite{knmi}, employing a typical inland location in the center of the Netherlands.
Economic data encompass forecasted day-ahead electricity prices coupled with residential natural gas prices. 
The former are collected and published by ENTSO-E (European Network of Transmission System Operators for Electricity) \cite{entsoe} and the latter from Eurostat \cite{eurostats} at granularities of 1 hour and 6 months respectively.
Both prices are coupled with environmental taxes set by the Dutch government \cite{taxdutch}, and electricity day-ahead prices additionally include fixed distribution and transmission tariffs from corresponding time periods to approximate best the end-user total price. These are reported by the Netherlands authority for consumers and markets, in Annex 2 \cite{acmelec}.

\begin{figure}
    \centering
    \begin{subfigure}[b]{0.95\linewidth}
       \includegraphics[width=.95\linewidth]{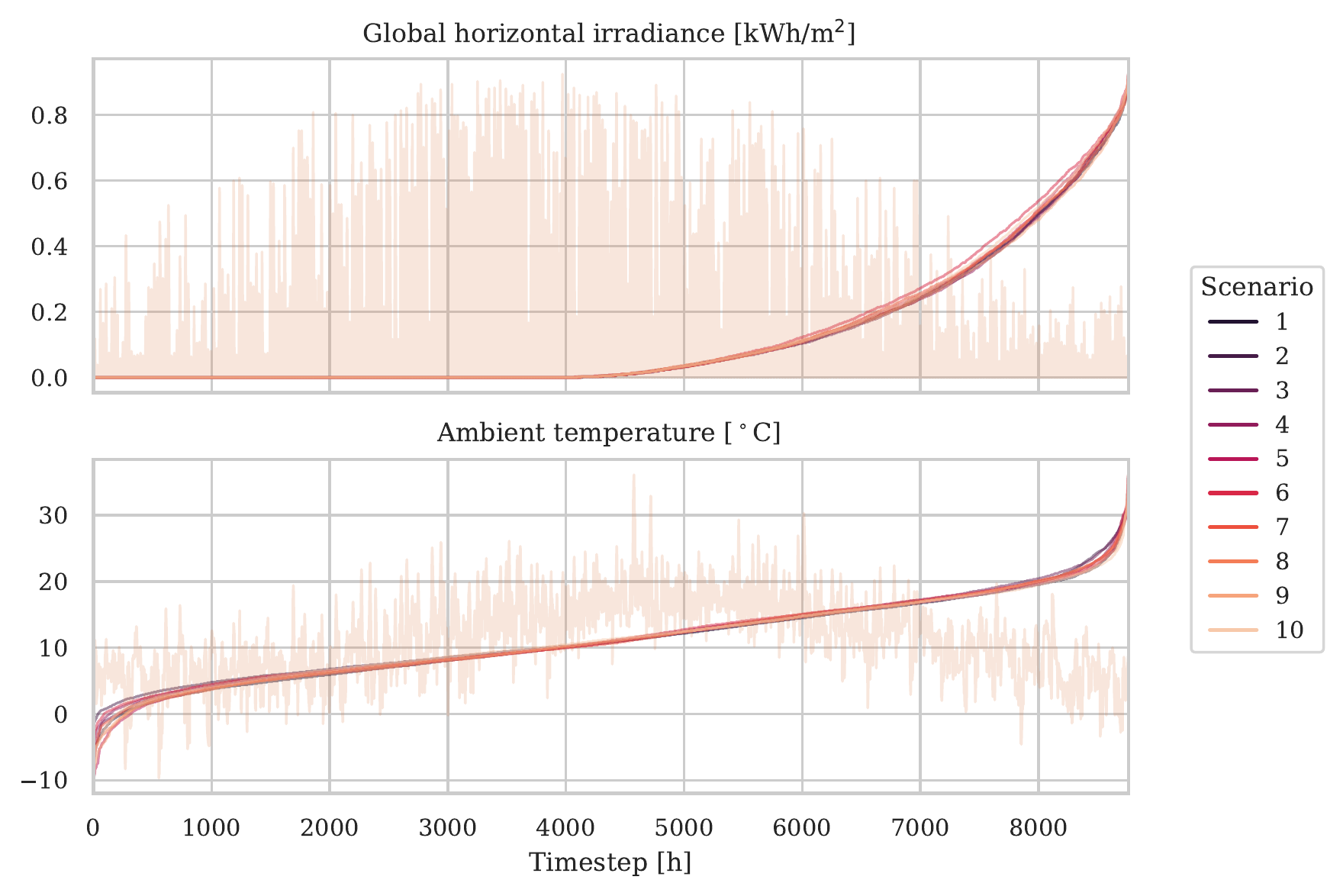}
       \caption{Weather conditions}
       \label{fig:loadweather} 
    \end{subfigure}
    
    \begin{subfigure}[b]{0.95\linewidth}
       \includegraphics[width=.95\linewidth]{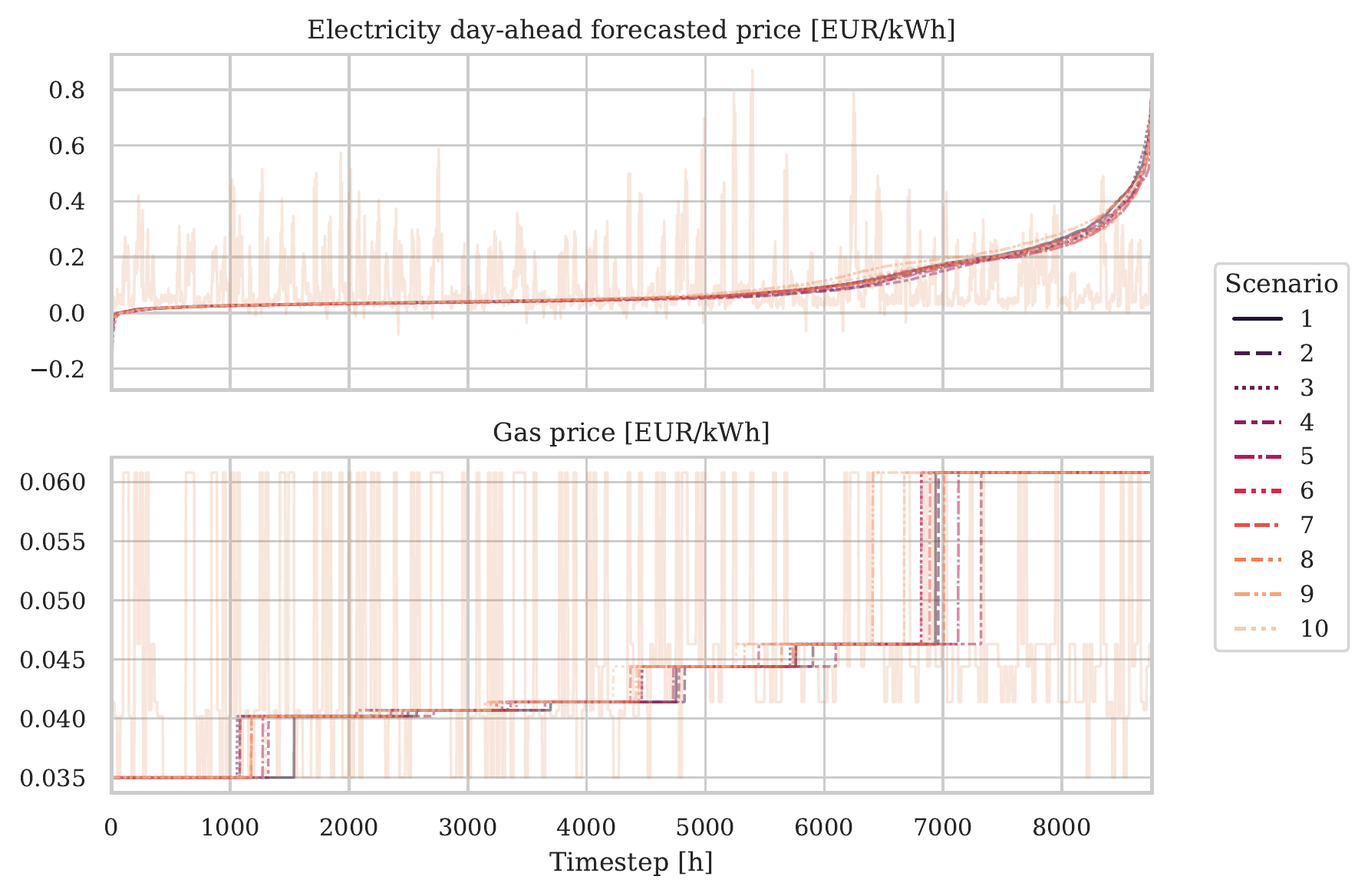}
       \caption{Economic circumstances}
       \label{fig:loadeco}
    \end{subfigure}
    
    \begin{subfigure}[b]{0.95\linewidth}
       \includegraphics[width=.95\linewidth]{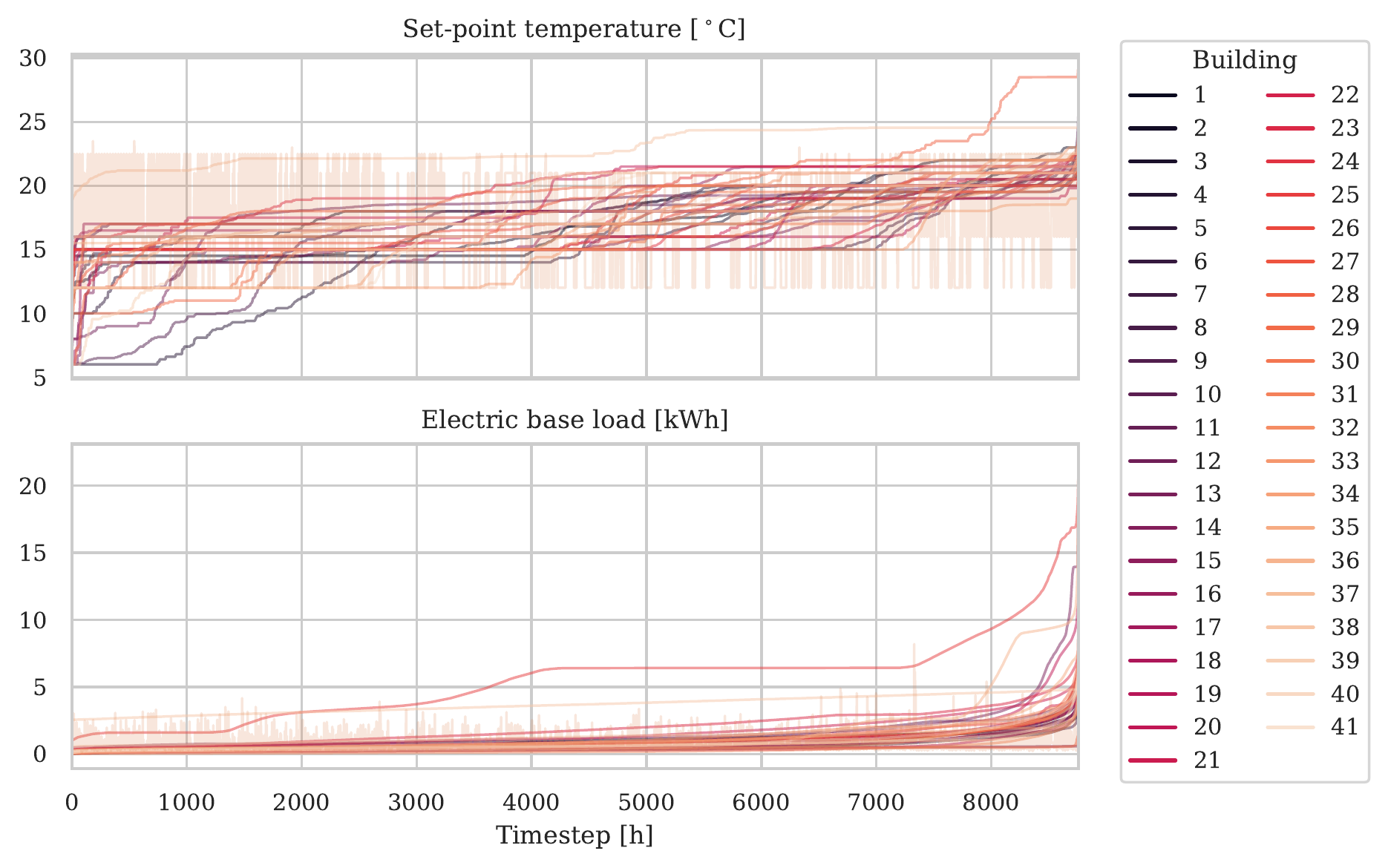}
       \caption{Occupant behavior averaged over representative scenarios}
       \label{fig:loadocc}
    \end{subfigure}
    \caption{Load curves of uncertainty factors per representative scenarios}
    \label{fig:load_curves}
\end{figure}

Load curves of identified representative scenarios are illustrated for weather, economic, and occupant behavior in Figure \ref{fig:load_curves}.

The techno-economic model parameters are gathered from the Danish Energy Agency (DEA) technology data catalogue \cite{deatechnodata}. 
Their referencing is summarized in Tab. \ref{tab:param}.

\begin{center}
	\begin{table*}
	\caption{Techno-economical parameter DEA catalogue referencing {\label{tab:param}}}
	\vspace{-0.5em}
	\setlength\extrarowheight{-3pt}
	\begin{adjustbox}{width=0.96\linewidth}
	    \begin{tabular}{ l c l c }
\toprule
Utility & Technology data catalogue & Index & Date \\
\midrule
EL & Energy Carrier Generation and Conversion June 2017 & 86 Hydrogen production via alkaline electrolysis (AEC) for 1MW plant & 2030 \\
FC & power and heat production plants & 12 Low temp PEM fuel cell - back pressure - hydrogen - small & 2020 \\
PV building & power and heat production plants & 22 rooftop PV residential & 2020 \\
PV community & power and heat production plants & 22 rooftop PV comm.\&industrial & 2020 \\
BOL & heating installations & 202 Gas boiler, ex single & 2020 \\
HP  & heating installations & 207 Heat pump, Air-to-water - apartment complex - existing building & 2020 \\
STC & heating installations & 215 Solar heating system - single-family house - existing building & 2020 \\
HWT & Energy storage & 142 Small-Scale Hot Water Tanks & 2020 \\
HYD & Energy storage & 151a Pressurized hydrogen gas storage system (Compressor \& Type I tanks \@ 200bar) & 2020 \\
BAT building & Energy storage & 181 Lithium-ion NMC battery (Utility-scale, Samsung SDI E3-R135) & 2020 \\
BAT community & Energy storage & 182 NaS battery & 2020 \\
\bottomrule
\end{tabular}
	\end{adjustbox}
	\end{table*}
\end{center}

    \section{Results and Discussion}\label{sect_res}
The identified optimal energy community design is here presented along with its in situ building control operational strategy. 
Then we unveil which uncertainty factor impacts the community design most through a local sensitivity analysis.

\subsection{Resilient energy community design}
\begin{figure}
    \centering
    \begin{adjustbox}{width=0.99\linewidth}
    \includegraphics{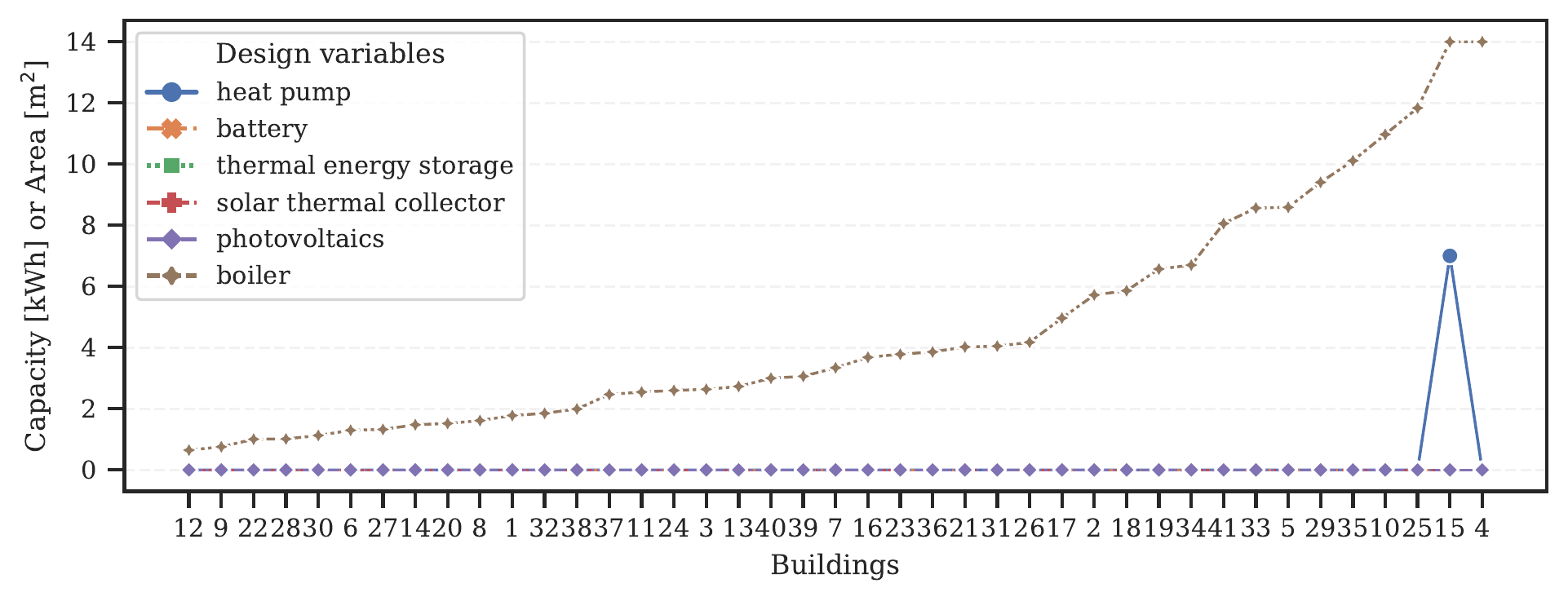}
    \end{adjustbox}
    \caption{Optimal design of the energy community by taking stochasticity into account.}
    \label{fig:stochastic_optimal_design}
\end{figure}
\begin{figure}
    \centering
    \begin{adjustbox}{width=0.99\linewidth}
    \includegraphics{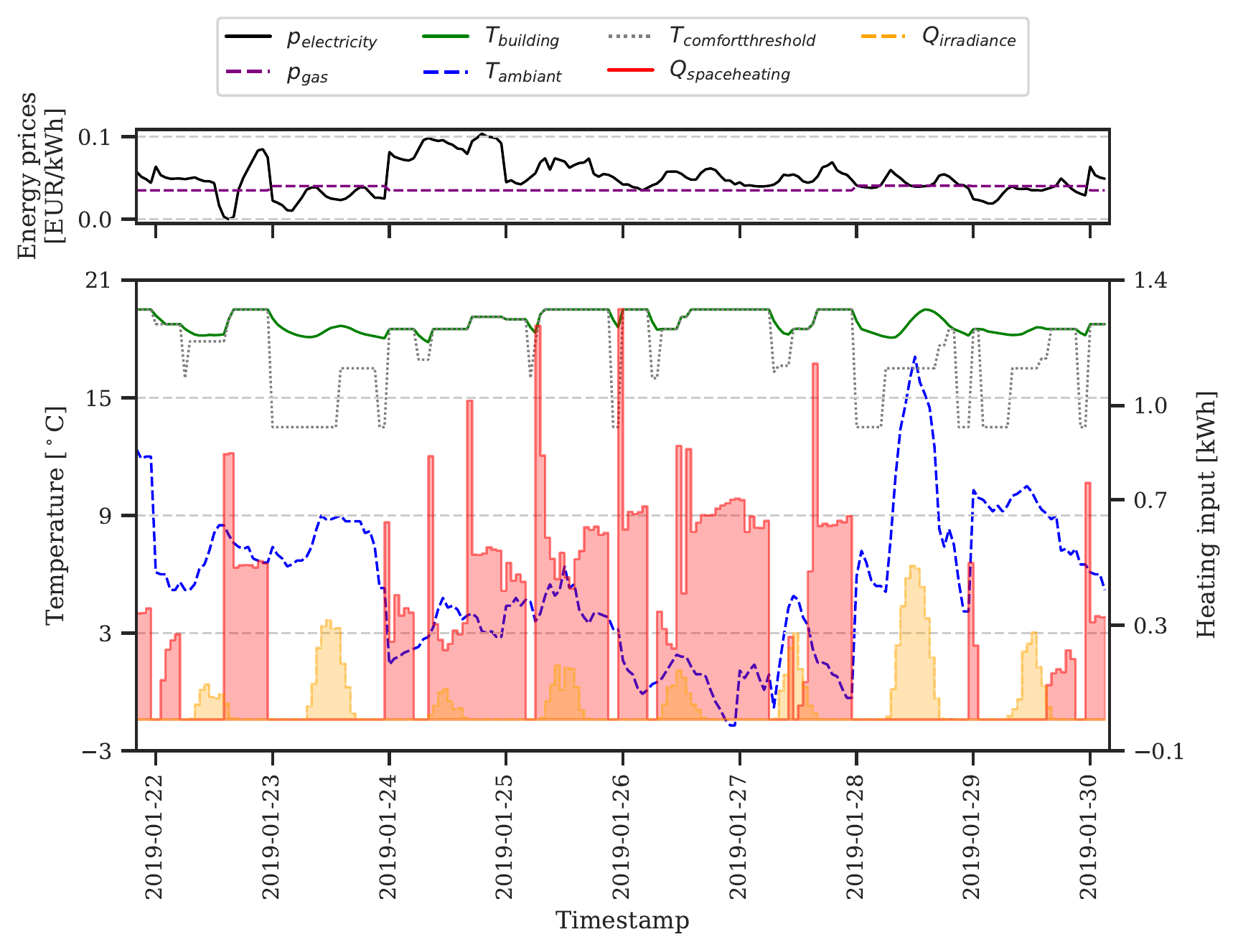}
    \end{adjustbox}
    \caption{Building \#1 inside temperature control.}
    \label{fig:blg_control_all}
\end{figure}

The distributed stochastic optimization problem was found to converge immediately following its first iteration as no improvements to the master problem objective function were made in the following iterations.
This direct convergence is a result of the identified optimal design of the energy community, which solely considers the necessary utilities to provide heating to the buildings, see Figure \ref{fig:stochastic_optimal_design}. 
The optimal design variables selected display boilers as the most energy and cost-efficient utility to supply space heating. Building \#15 is the only system considering a heat pump to supplement its heating needs, due to the maximum capacity reached by the boiler. Another building, i.e., building \#4, also reaches maximum boiler capacity, however, does not consider the additional input of a heat pump to provide its space heating needs.

This optimal energy community design renders energy exchanges between the different building systems of the community disadvantageous thus resulting in a situation where cooperation between energy community members is not exploited.
The efficiency and investment costs of boilers coupled with the lower energy prices of gas, make heat pumps an unprofitable alternative for space heating and here thus only considered as a complementary utility.
Higher gas prices coupled with carbon reduction incentives are consequently still needed for buildings to consider heat pumps as an interesting alternative to boilers and begin the decarbonization of the sector.

As gas prices are commonly fixed within fixed tranches of months, demand-side management control strategies in response to varying energy prices are seldom observed in the building in situ control strategies.
Figure \ref{fig:blg_control_all} illustrates the inside temperature control of building \#1 in function of occupant-driven comfort requirements, ambient and economic conditions.
The comfort requirements defined by the occupants can, however, be observed to be set at lower plateaus in periods of higher gas prices, see days 01/23 and 01/28 where the set-point temperature is set at a higher plateau of 17$^\circ$C instead of the more common 19$^\circ$C.

\subsection{Uncertainty factor impact assessment}
\begin{figure}
    \centering
    \begin{adjustbox}{width=0.99\linewidth}
    \includegraphics{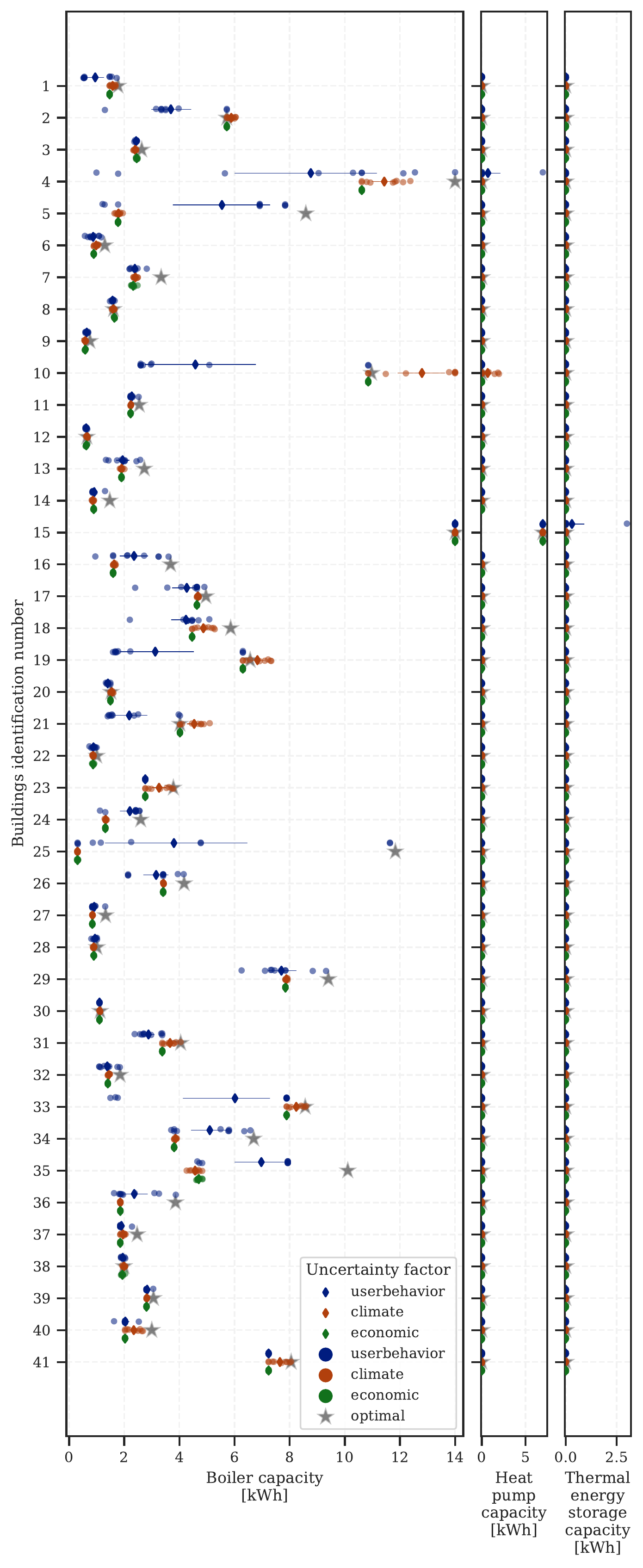}
    \end{adjustbox}
    \caption{Uncertainty factor impact on energy community design.}
    \label{fig:SA_allheatingutilities}
\end{figure}

To quantify the impact of occupant behavior on energy community planning within the context of other uncertainty factors we evaluate the optimal design of considered utilities within varying uncertain parameters.
Figure \ref{fig:SA_allheatingutilities} presents the spread and mean values of the considered design variable over the building stock, i.e., full circle and diamond points respectively, grouped by uncertainty parameter. Full lines represent the standard deviation of design values per uncertainty factor.
The optimal design values accounting for all uncertain parameters, i.e., presented in Fig. \ref{fig:stochastic_optimal_design}, are here highlighted as \emph{optimal} to showcase the optimal design relative to the uncertainty affecting the energy system.
The spread of values exposes occupant behavior as the factor with the largest impact on selected design variables, followed by climate and economic conditions.
Although climate conditions often produce design variable spreads with higher central tendencies, i.e., engendering larger design variables on average, the dispersion of values is highest for occupant behavior. 
Buildings \#4 and \#10 are two examples where heat pumps are considered only when sensitive to particular uncertainty factors, namely, occupant behavior and climate conditions respectively. In both cases, the central tendency of boiler capacities is highest for climate-related uncertainties but their spread is highest for occupant-driven ones. The main difference being that identified boiler capacities of building \#4  range from 1 to 14 kWh and require the additional heat pump investment in one setting while for building \#10 occupant-driven uncertainties result in much lower boiler capacities than climate-related ones subsequently resulting in climate-driven heat pump investment in 4 scenarios.
Building \#15 on the other hand necessitates maximum design capacities both for boiler and heat pumps for all uncertainty-related factors resulting in its need for thermal storage investment in one occupant-driven setting. This is likely due to large set-point temperature shifts set by the occupant requiring additional heat inputs in specific time windows.
This important finding demonstrates the significance of occupant behavior in strategic urban energy design and the value of bridging these two disconnected spatial scales.

Furthermore, optimal design variables identified by the stochastic problem formulation can often be found with values higher than its highest uncertainty analysis factor, e.g., see buildings \#5, 7, 25, or 35. 
This is a result of the separation of correlated uncertainty factors in the sensitivity analysis. 
Indeed, in the local sensitivity analysis, evaluated uncertainty factors will be paired to nominal scenarios of other factors, e.g., occupant behavior scenarios $\omega_{i}^{occ}$ will be paired to $\omega_{nom}^{eco}$ and $\omega_{nom}^{clim}$, whereas in the stochastic problem formulation, the combinations are different and each scenario $i$ regroups all uncertainty factors.
As nominal scenarios represent the most likely scenario per uncertainty factor, it is likely other scenarios might impact the design with more unlikely, and possibly extreme conditions, thus resulting in larger design variables.
This highlights the importance of considering varying uncertainty parameters in the design phase of energy systems and the value brought by stochastic approaches, which provide robust solutions towards the more extreme conditions.

    \section{Conclusion}\label{sect_con}

This paper attempts to bridge two typically disconnected scales of the built environment for improved energy and carbon emission performances: occupants and the urban energy system.
Strategic energy planning is undertaken by exploiting energy community concepts such as peer-to-peer cooperative energy exchanges and shared neighborhood-level infrastructure.
Particularly, uncertainty factors affecting urban energy planning are embedded to the problem and investigated by employing a stochastic problem formulation supplemented by a local sensitivity analysis.
Computational tractability concerns are addressed, founded on an organic spatial problem distribution, which we validate by a proof of concept.
The setup notably echoes that of decentralized energy management systems, thus implanting our approach in a real-world operational control setting, suitable for field deployment.

From historical measurements and accurate techno-economical parameter settings, a typical Dutch energy community composed of 41 residential buildings is designed.
Results present a fast-converging distributed stochastic problem, where boilers are showcased as the winning utility provider for space heating. 
These expose current Dutch energy prices along with carbon emission taxes as not profitable enough for generalized heat pump adoption in typical residential buildings. 
It is postulated that increased electricity prices might also push energy communities to further adopt distributed energy renewables such as photovoltaics and solar thermal collectors. In such as setting, energy storage utilities will become compelling to align mismatches between renewable production and occupant-driven energy loads, as well as peer-to-peer energy exchanges.

The impact of occupant behavior, encompassing set-point temperature and smart-meter base loads, on strategic energy planning is specifically investigated relative to other uncertainty factors, i.e., economic, electricity and gas prices, and weather, ambient temperature and solar irradiance, conditions.
The analysis reveals occupants to be the leading factor affecting energy community design, thus confirming the relevance of our approach in connecting occupants to urban energy planning.

Lastly, all the implementations of this paper are open-sourced to encourage research dissemination and favor research reproducibility\footnote{\href{https://github.com/JulienLeprince/energycommunityplanning}{https://github.com/JulienLeprince/energycommunityplanning}}.

\subsection{Limitations and future research}
While our findings portray occupant behavior, i.e., building set-point temperature and electricity base loads, to be the leading uncertainty factor affecting the system design, it should be noted that representative scenarios were only sampled from historical measurements over the years 2019 to 2022.
Employing older historical measurements or considerations with regard to the long-term evolution of weather and economic data might produce differing results.
Thus, forecasts and uncertainty analysis related to these developments remain a goal for future research.

Additionally, varying community sizes and heterogeneity, i.e., number of buildings and representative occupant behaviors respectively, in the context of optimal stochastic urban energy planning offers an interesting analysis for urban planning decision-makers. 
Answering questions such as "\textit{How large must a community be for shared utilities, such as seasonal storage, to become profitable?}" or "\textit{How does occupant heterogeneity affect energy saving potentials?}" provide appealing research interrogations to guide subsequent studies.

\section{CRediT authorship contribution statement}\label{sect_credit}

\textbf{Julien Leprince}: Conceptualization, Methodology, Software, Data curation, Formal analysis, Visualization, Writing - original draft, review and editing.
\textbf{Amos Schledorn}: Conceptualization, Methodology, Software, Formal analysis, Writing - review and editing.
\textbf{Daniela Guericke}: Methodology, Supervision, Validation, Writing - review and editing.
\textbf{Dominik Franjo Dominkovic}: Methodology, Supervision, Validation, Writing - review and editing.
\textbf{Henrik Madsen}: Methodology, Supervision, Validation, Writing - review and editing.
\textbf{Wim Zeiler}: Supervision, Funding acquisition.

All authors have read and agreed to the published version of the manuscript.

\section{Acknowledgments}
This work is funded by the Dutch Research Council (NWO), in the context of the call for Energy System Integration \& Big Data (ESI-bida). Support by SEM4Cities funded by Innovation Fund Denmark (Project No. 0143-0004) is also gratefully acknowledged. 
Finally, we wish to express our appreciation to Eneco, with particular thanks to Dr. Kaustav Basu and Rik van der Vlist, as well as Eneco customers for their contributions to this research.

\bibliography{mybibfile}

\end{document}